\documentclass[aps, prx, reprint, superscriptaddress, longbibliography, floatfix]{revtex4-2}
\usepackage{graphicx}% Include figure files
\usepackage{placeins}
\usepackage{dcolumn}% Align table columns on decimal point
\usepackage{bm}% bold math
\usepackage{physics}
\DeclareMathAlphabet{\pazocal}{OMS}{zplm}{m}{n}
\usepackage{amsmath}
\usepackage{amsfonts}
\usepackage{mathrsfs}
\usepackage{subfigure}
\usepackage{color,soul}
\usepackage{xcolor}
\usepackage[colorlinks=true,linkcolor=blue,anchorcolor=red,citecolor=blue, urlcolor=blue]{hyperref}

\begin{document}
\preprint{APS/123-QED}
 
\title{Recurrent convolutional neural networks for modeling non-adiabatic dynamics of quantum-classical systems}

\author {Alex Ning}
\affiliation{Department of Computer Science, University of Virginia, Charlottesville, Virginia, 22904, USA}
\affiliation{Department of Mathematics, University of Virginia, Charlottesville, Virginia, 22904, USA}

\author {Lingyu Yang}
\affiliation{Department of Physics, University of Virginia, Charlottesville, Virginia, 22904, USA}

\author {Gia-Wei Chern}
\affiliation{Department of Physics, University of Virginia, Charlottesville, Virginia, 22904, USA}

\begin{abstract}{
Recurrent neural networks (RNNs) have recently been extensively applied to model the time-evolution in fluid dynamics, weather predictions, and even chaotic systems thanks to their ability to capture temporal dependencies and sequential patterns in data. Here we present a RNN model based on convolutional neural networks for modeling the nonlinear non-adiabatic dynamics of hybrid quantum-classical systems. The dynamical evolution of the hybrid systems is governed by equations of motion for classical degrees of freedom and von Neumann equation for electrons. The physics-aware recurrent convolution (PARC) neural network structure incorporates a differentiator-integrator architecture that inductively models the spatiotemporal dynamics of generic physical systems. We apply our RNN approach to learn the space-time evolution of a one-dimensional semi-classical Holstein model after an interaction quench. For shallow quenches (small changes in electron-lattice coupling), the deterministic dynamics can be accurately captured using a single-CNN-based recurrent network. In contrast, deep quenches induce chaotic evolution, making long-term trajectory prediction significantly more challenging. Nonetheless, we demonstrate that the PARC-CNN architecture can effectively learn the statistical climate of the Holstein model under deep-quench conditions.
}

\end{abstract}
\date{\today}
\maketitle

\section{Introduction}

\label{sec:intro}

Recurrent Neural Networks (RNNs) are an established framework for analyzing sequential or dynamic data with applications across physics~\cite{hochreiter97,graves12,lipton15,sherstinsky20}. Unlike feed-forward networks, which assume independence between inputs, RNNs incorporate internal feedback loops which can capture dependencies between sequential data. This is a crucial behavior when modeling the evolution of time-dependent physical systems and trajectories with strong temporal correlations between data points. RNNs can learn the underlying dynamics from time-series data without requiring explicit knowledge of governing equations. They can predict future states of a system based on its past behavior. Indeed, RNNs are among one of the prominent deep-learning approaches often employed to solve partial differential equations (PDEs)~\cite{hu20,karlbauer22,hafiz24,wu22,sun20,zhong21,liang24,long18,long19,zuo15,saha21,ren22,nguyen23,nguyen24}, which are the ubiquitous tools for modeling the spatiotemporal evolution of most physical systems.

The essential idea behind the RNN approach to PDE, namely using current and previous step states to calculate the state at the next time step, is actually similar to most time-stepping numerical methods for solving time-dependent PDEs, such as Euler’s, Crank-Nicolson, and  higher-order Runge-Kutta schemes~\cite{press92,morton05,ascher97}. The RNN serves as a surrogate time-stepping method replacing the numerical finite-difference or finite-element schemes. The numerical solution on each of the grid point (for finite difference) or grid cell (for finite element) computed at a set of contiguous time points can be treated as neural network input in the form of one time sequence of data. The deep learning framework is then trained to predict any time-dependent PDEs from the time series data supported on the grids. In other words, the supervised training process can be viewed as a way for the deep learning framework to learn the numerical solution from the input data by optimizing the coefficients of the neural network layers.

A central theme in recent research - particularly motivated by applications to chaotic dynamics in many physical systems - is learning the long-term climate of chaotic systems, characterized by ergodic statistics such as attractor geometry and Lyapunov spectra, rather than insisting on accurate long-horizon trajectories for individual initial conditions~\cite{Pathak_2017,Haluszczynski_2019}. For canonical low- and moderate-dimensional benchmarks, as well as spatiotemporal PDEs, echo state networks have been trained to reproduce climate measures - including Lyapunov spectra - for systems such as Lorenz and the Kuramoto-Sivashinsky equation, with scalability to large spatial domains via parallel reservoir architectures~\cite{Pathak_2017,PhysRevLett.120.024102}. Robustness analyses across multiple training realizations and network topologies reveal substantial variability, with runs achieving better short-term accuracy also tending to reproduce the long-term climate more faithfully~\cite{Haluszczynski_2019}. Similarly, in transitional shear flows, models trained exclusively on turbulent data can nevertheless capture laminar behavior and the statistics of turbulent-laminar switching~\cite{PhysRevE.107.014304}. This emphasis on statistical fidelity spans diverse implementations, from large parallel reservoirs to minimal physical reservoirs - such as a single driven pendulum - capable of performing both temporal and non-temporal tasks by exploiting transient dynamics~\cite{PhysRevE.105.054203}.

The dynamics of physical systems is also characterized by the emergence of complex spatial patterns, such as stripes and vortices. This indicates the importance of spatiotemporal correlations in the dynamical modeling of such physical systems. Yet, traditional neural networks often perform poorly at capturing local relationships due to their fully-connected nature; Convolutional Neural Networks (CNNs)~\cite{goodfellow16}, on the other hand, excel at such tasks. CNNs are a variety of neural network which utilize \textit{convolution} layers to process a grid or array of values using a ``sliding" \textit{kernel} to extract learned features with geometric structure. Convolutional layers strongly exploit patterns with geometric locality in inputs, and within a CNN many convolutional layers are often composed to extract a hierarchy of features starting from low-level patterns to high-level and complex relationships. Because of these unique capabilities, CNNs have become a crucial component in deep-learning approaches to solving PDEs~\cite{saha21,ren22,nguyen23,nguyen24}. 

A recently proposed RNN scheme, dubbed physics-aware recurrent convolution (PARC)~\cite{nguyen23,nguyen24}, further incorporates physics information into its architecture, thus offering improved accuracy and representation capability. Indeed, machine learning (ML) models with physical constraints explicitly incorporated have attracted great interest among researchers. For example, the ML force-field framework for {\em ab initio} molecular dynamics utilizes the locality principle to implement  local energy or force models, enabling linear scaling computation~\cite{behler07,bartok10,li15,shapeev16,botu17,chmiela17,chmiela18}. Another example is the equivariant neural networks which allows physical symmetries to be explicitly included on a supervised learning model. Their architecture is designed in such a way that the neurons of each layer exhibit well-defined transformation properties under operation of a given symmetry group~\cite{weiler18,batzner22,gong23,ma23}. 

Compared with other RNNs, PARC is structured as two separate but directly connected CNN-based differentiator and integrator components. The two-stage architecture is motivated by algorithms of most numerical methods for PDEs: the derivatives of the field $d\mathbf u / dt$ are first computed from the current configuration, which is then integrated to produce the future states of the evolving field. The stand-alone integral solver also offers the capability for stable, longer time predictions. This differentiator-integrator structure of PARC thus provides a prior on integration tasks, resulting in its ``physics-aware" nature and enabling improved data efficiency and performance.

In this work we propose a deep-learning approach based on both a standard CNN and the PARC architecture for modeling the non-adiabatic dynamics of lattice models with coexisting classical and quantum electron degrees of freedom. Such hybrid quantum-classical models are an important simplification for modeling realistic physical systems with, e.g. multiple length and time scales. One of the most prominent examples is the {\em ab initio} molecular dynamics methods widely used in quantum chemistry and materials science~\cite{marx09}. Yet, direct dynamical simulation of such hybrid systems is still a challenging computational task because of the exponentially large dimension of the Hilbert space associated with the quantum subsystem. 

Even in the absence of direct electron-electron interactions, a many-body description of  electrons is still required in order to accommodate the Fermi-Dirac quantum statistics. In the non-interacting case, the complexity of simulating a many-electron wave function can be reduced to the dynamics of single-electron density matrix or correlation functions $\rho(\mathbf r, \mathbf r', t)$ where $\mathbf r$, $\mathbf r'$ are coordinates of lattice sites. As a result, for a $D$-dimensional system $\mathbf r \in \mathbb{Z}^D$, the minimum description in terms of density matrix requires $D^2$-dimensional time-dependent differential equations. As a proof-of-principle, here we consider the 1-dimensional (1D) semi-classical Holstein model and demonstrate a CNN-based framework for its non-adiabatic dynamics under a quantum quench. 

Although the dynamics of discrete lattice systems is formally described by coupled ordinary differential equations (ODEs), spatial couplings between dynamical variables nonetheless resemble those of discretized PDEs using either finite-difference or finite-element schemes. The spatial correlations and potential emergent spatial patterns can be efficiently captured by a CNN, as discussed above. Indeed, similar CNN-based force-field models have been developed for spin dynamics of the so-called s-d system, a lattice model of metallic magnets~\cite{cheng23b}. Importantly, the convolution operation with a finite-sized kernel naturally incorporates the locality principle, which can be straightforwardly scaled to larger systems.

Before closing the Introduction section, we note that PARC relies on the inductive bias method~\cite{karniadakis21} to embed prior physics knowledge within the neural network structures. This is in contrast to the Physics-Informed Neural Networks (PINNs)~\cite{raissi19}, a popular deep-learning architecture for solving PDEs, which are based on learning-biased approach where physical constraints are directly enforced by minimizing the PDE residues and boundary/initial conditions through a loss function. Although PINNs can achieve high performance with no training data besides the initial condition, a single PINN is in general only capable of approximating the solution to a single initial condition, making it unfit for the generalized task of integration on a family of solutions.

The remainder of the paper is organized as follows. Section~\ref{sec:holstein} introduces the non-adiabatic dynamics of the semi-classical Holstein model within the Ehrenfest dynamics framework. In Section~\ref{sec:standard_CNN}, we describe the formulation, implementation, and results of a standard CNN model applied to the Holstein model under a shallow quench, demonstrating that even a compact CNN can accurately capture the time evolution. Section~\ref{sec:PARC_CNN} presents the formulation and implementation of a PARC-based model for the Holstein model under a deep quench, with results and benchmark analyses discussed in Section~\ref{sec:deep_results}. There we show that the trained model successfully reproduces the statistical climate of the trajectories, as evidenced by the agreement of the autocorrelation function. Finally, Section~\ref{sec:summary} provides a summary and outlook. All source code and model weights are available at~\href{https://github.com/apning/holstein-parc}{\texttt{github.com/apning/holstein-parc}}.

\section{Non-adiabatic dynamics of Holstein model}

\label{sec:holstein}

We consider a 1D Holstein model~\cite{holstein59} with spinless fermions, described by the following Hamiltonian with three parts
\begin{eqnarray}
	& & \hat{\mathcal{H}} = - t_{\rm nn} \sum_{i} \left(\hat{c}^{\dag}_{i} \hat{c}^{\,}_{i+1} + \hat{c}^{\dag}_{i+1} \hat{c}^{\,}_{i} \right)
	-g\sum_{i}\left( \hat{n}_i -\frac{1}{2}\right) \hat{Q}_{i} \nonumber \\
	& & \qquad + \sum_{i}\left( \frac{1}{2m} \hat{P}^{2}_{i}+\frac{1}{2}m\Omega^2 \hat{Q}^{2}_{i} \right).
\end{eqnarray}
Here $\hat{c}^\dagger_i$ ($\hat{c}^{\,}_i$) denotes the creation (annihilation) operator of an electron at lattice site-$i$, $Q_i$ represents a local phonon degree of freedom, $P_i$ is the associated conjugate momentum, $m$ is an effective mass, $\Omega$ is the intrinsic oscillation frequency, and $K\equiv m\Omega^2$ is the force constant, finally $g$ denotes the electron-phonon coupling coefficient. The first term $\hat{\mathcal{H}}_{\mathrm{e}}$ describes the hopping of electrons between nearest neighboring sites. The second part essentially describes the Einstein phonon model, which corresponds to a set of simple harmonic oscillators each associated with a lattice site. The third term describes a local coupling between the electron density $\hat{n}_i = \hat{c}^\dagger_i \hat{c}^{\,}_i$ and the displacement of the oscillator. 

The Holstein model at half-filling on various bipartite lattices exhibits a robust commensurate charge density wave (CDW) order that breaks the sublattice symmetry~\cite{Noack91,Zhang19,Chen19,Esterlis19}. In one dimension, this commensurate CDW order is characterized by a ultra-short period modulation of electron density, 
\begin{eqnarray}
    \label{eq:n-mod}
	\langle \hat{n}_i \rangle = \overline{n} + \delta n \cos\left(\frac{\pi}{a} x_i \right), 
\end{eqnarray}
where $\langle \cdots \rangle$ denotes ground-state expectation value, $\overline{n} = 1/2$ is the average density, $\delta n$ is the modulation amplitude, $a$ is the lattice constant, and $x_i = x_0 + i \, a$ is the physical coordinate of site-$i$.  It is worth noting that the CDW order remains robust even in the semi-classical approximation. Indeed, the semi-classical phase diagram of the CDW order obtained by a hybrid Monte Carlo method agrees very well with that obtained from determinant quantum Monte Carlo simulations~\cite{Esterlis19}. Within the semi-classical approximation for the CDW dynamics, the Holstein is an example of the hybrid quantum-classical systems discussed above.

Here we employ the Ehrenfest dynamics framework~\cite{li05x,marx09} to describe the semi-classical dynamics of the Holstein model. A similar semi-classical dynamics method was recently employed to study the photo-emission and long-time behaviors of CDW states in the 1D Holstein model~\cite{Petrovic22}.  To this end, we assume a product form for the quantum state of the system: $\ket{\Gamma(t)}=\ket{\Phi(t)} \otimes \ket{\Psi(t)}$, where $\ket{\Phi(t)}$ and $\ket{\Psi(t)}$ denote the phonon and electron wave-functions, respectively. The semi-classical approximation for the lattice subsystem amounts to a direct product wave function $\ket{\Phi(t)}=\prod_{i} \ket{\phi_{i}(t)}$ for the phonons. As a result, the expectation value of phonon operators, e.g. $\langle \Gamma(t) | \hat{Q}_i | \Gamma(t) \rangle$ reduces to $Q_i(t) \equiv \langle \phi_i(t) | \hat{Q}_i | \phi_i(t) \rangle$, and similarly for the momentum operators, $P_i(t) \equiv \langle \Gamma(t) | \hat{P}_i | \Gamma(t)\rangle = \langle \phi_i(t) | \hat{P}_i |\phi_i(t) \rangle$. 

The dynamics of these `classical' variables is given by the expectation of Heisenberg equation, e.g. $d \langle \hat{P}_{i} \rangle/dt = -i \langle [ \hat{P}_{i}, \hat{\mathcal{H}} ] \rangle / \hbar$, where $\langle ... \rangle$ is the expectation value computed using the full wave-function $\ket{\Gamma(t)}$. Direct calculation of the commutators yields the coupled Hamiltonian dynamics
\begin{eqnarray}
	\label{eq:newton_eq}
	\frac{dQ_i}{dt} = \frac{P_i}{m}, \qquad \frac{dP_i}{dt} = g n_i - K Q_i.
\end{eqnarray}
Here the time-dependent electron density is given by $n_i(t) = \langle \Gamma(t) | \hat{n}_i | \Gamma(t) \rangle$, which can be simplified to $n_i(t) = \langle \Psi(t) | \hat{n}_i | \Psi(t)\rangle$ thanks to the product form of the system quantum state.

Since the Holstein Hamiltonian is quadratic in electron operators, the time evolution of the many-electron Slater-determinant wave function can be exactly solved numerically. A more efficient approach is based on the dynamical equation for the single-particle density matrix, 
\begin{eqnarray}
	\label{eq:rho1}
	\rho_{ij}(t) = \bra{\Psi(t)} \hat{c}^\dagger_j \hat{c}^{\,}_{i} \ket{\Psi(t)}.
\end{eqnarray}
The on-site electron number, which is a driving force of the lattice dynamics in Eq.~(\ref{eq:newton_eq}), is readily given by the diagonal elements: $n_i(t) = \rho_{ii}(t)$. The dynamical evolution of the density matrix is governed by the von~Neumann equation 
\begin{eqnarray}
	\label{eq:drho_dt}
	\frac{d\rho}{ dt} = -i [H(\{Q_i\}), \rho ], 
\end{eqnarray}	
where $H$ is a matrix which can be viewed as the first-quantized Hamiltonian on a lattice. Explicitly, $H_{ij} = -t_{nn} (\delta_{j, i+1} + \delta_{j, i-1} )- g\delta_{ij} Q_i(t)$. Direct calculation gives
\begin{eqnarray}
\label{eq:von-neumann}
	\label{eq:von-neumann_eq}
    i\hbar \frac{d\rho_{ij}}{dt} =  \sum_{k}\left( \rho_{ik}t_{kj} - t_{ik}\rho_{kj} \right) + g\left( Q_{j} - Q_{i} \right)\rho_{ij}. \quad
\end{eqnarray}
There are two characteristic time scales for the dynamics of the Holstein model. First, from the bandwidth of the electron tight-binding model $W = 4 t_{\rm nn}$, one can define a time scale $\tau_{\rm e} = \hbar/ t_{\rm nn}$ for the electron dynamics. Another time scale is given by the natural frequency $\Omega$ of the local simple harmonic oscillator: $\tau_{\rm L} = 1/ \Omega$. The dimensionless adiabatic parameter is defined as the ratio $r = \tau_{\rm e} / \tau_{\rm L} = \hbar \Omega / t_{\rm nn}$.   The electron-phonon coupling can also be characterized by a dimensionless parameter $\lambda$ as follows. First, let $Q^*$ be lattice distortions estimated from the balance of elastic energy and electron-phonon coupling: $K Q^{* 2} \sim g \langle n \rangle Q^*$. Assuming electron number $\langle n \rangle \sim 1$, we obtain $Q^* \sim g / K$. The dimensionless parameter $\lambda = g Q^* / W = g^2 / W K$ corresponds to the ratio of electron-phonon coupling to the bandwidth. For all simulations discussed below, these two dimensionless parameters are set to $r = 0.4$ and $\lambda=1$. The simulation time is measured in unit of $\tau_{\rm e}$, energies are measured in units of $t_{\rm nn}$, and the lattice distortion is expressed in terms of $Q^*$.

\begin{figure*}[t]
\includegraphics[width=1.99\columnwidth]{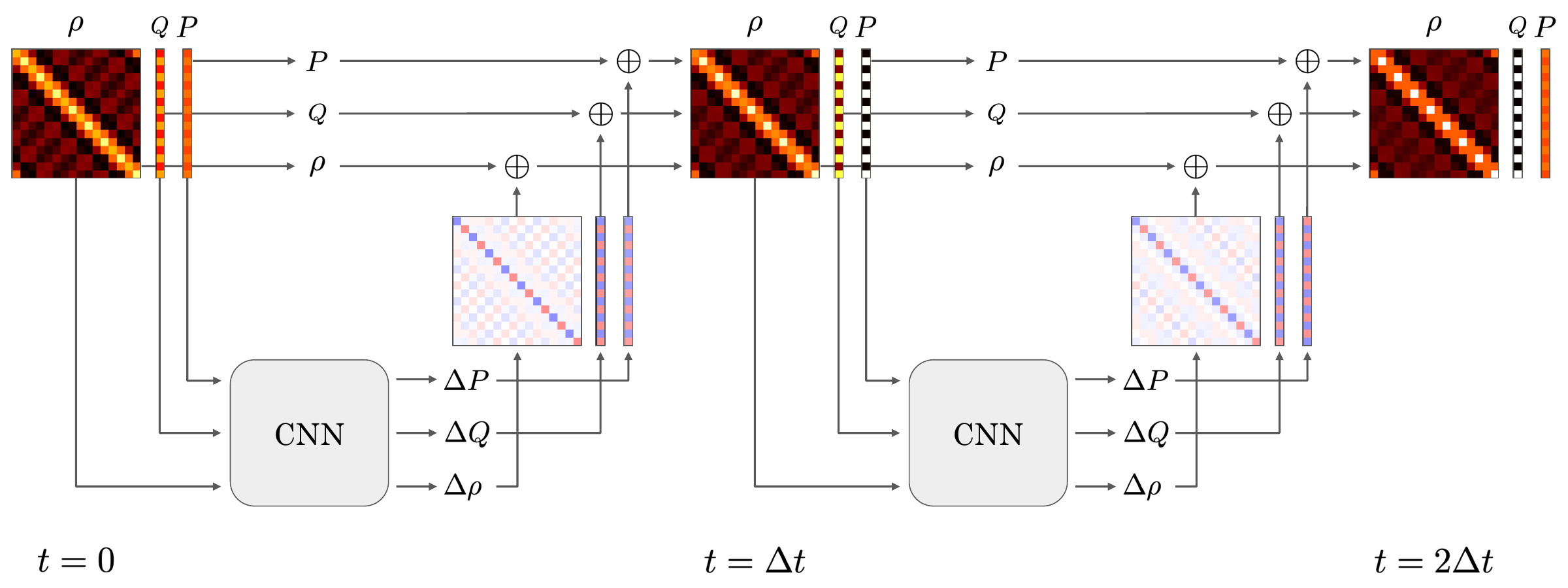}
\caption{\label{fig:standard_schematic}
Schematic diagram of the recurrent structure for non-adiabatic dynamics of the semi-classical Holstein model. The framework is based on a single CNN for all three dynamical degrees of freedom: $Q$, $P$, and $\rho$.
}
\end{figure*}

\begin{figure*}[t]
\includegraphics[width=1.55\columnwidth]{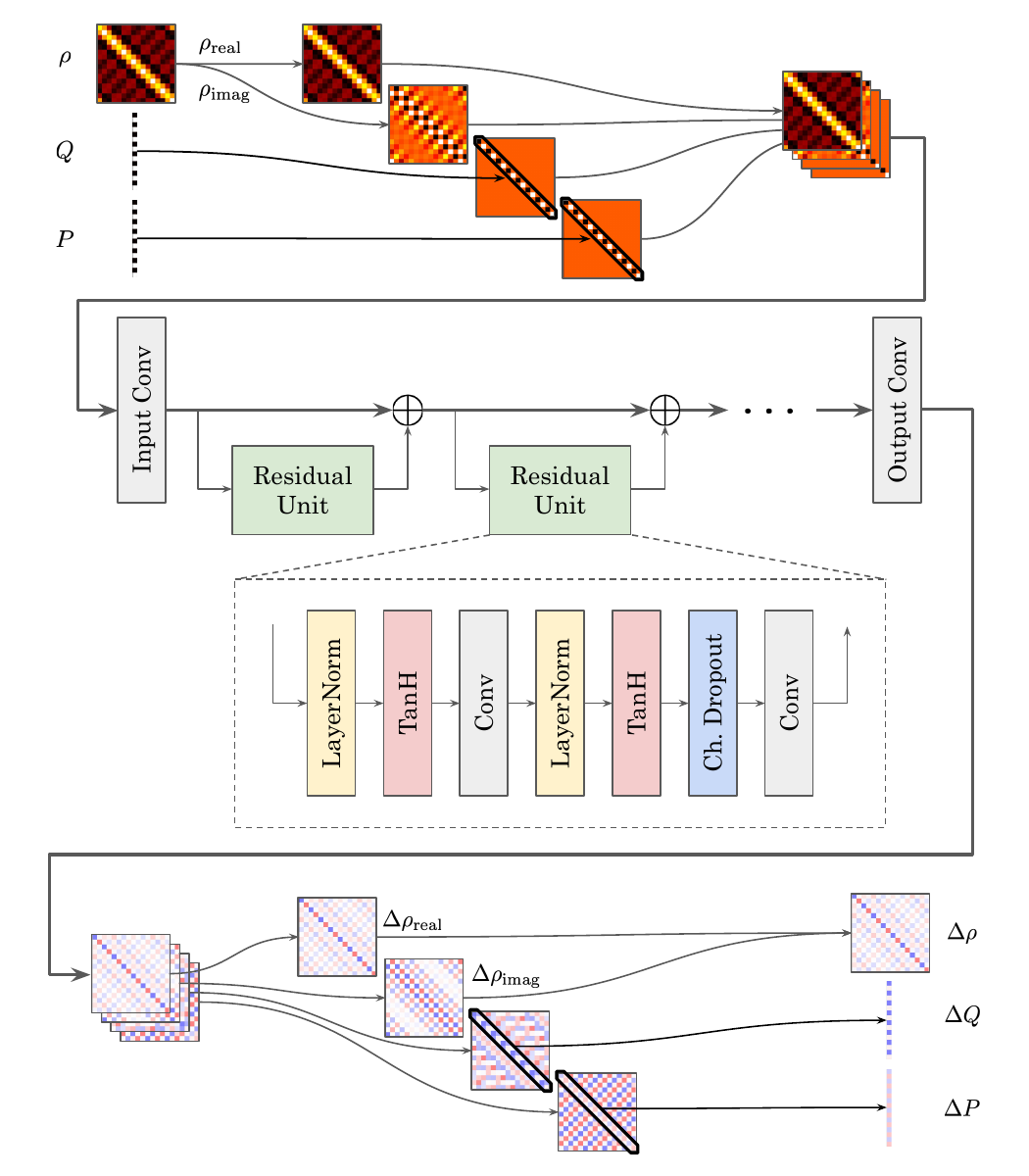}
\caption{\label{fig:standard_CNN_architecture}
Architecture of the standard CNN model. The components $\rho$, $Q$, and $P$ are first prepared for input. $\rho$ is split into real-valued components $\rho_\text{real}$ and $\rho_\text{imag}$. $Q$ and $P$ are both inserted into the diagonals of zero-matrices. The resulting $4$-channel grid is then input into the CNN model with modified ResNet-v2 based architecture. $\rho$, $Q$, and $P$ are then extracted from the output of the CNN in the reverse manner they were input.
}
\end{figure*}

\section{Standard CNN and Shallow Quench} \label{sec:standard_CNN}

In this section we consider RNN models for modeling the non-adiabatic dynamics of the 1D semi-classical Holstein model triggered by a small change in the electron-lattice coupling $g$, a scenario to be called shallow quench in the following. Our exact numerical simulations find that the resultant space-time evolution is less sensitive to noises in the initial conditions. Here we present a RNN scheme based on standard CNNs for modeling such relatively deterministic dynamics. While the state of the simple harmonic oscillators on a chain is described by two 1D arrays $\{ Q_i \}$ and $\{ P_i \}$, the minimum dynamical variables for the many-electron system are given by a density matrix $\rho_{ij}$, due to the quantum Fermi-Dirac statistics. This means that, for the simpler case of non-interacting electrons, the 1D quantum dynamics can be effectively modeled by an effective 2D classical dynamics.

\subsection{CNN-based recurrent structure}

For convenience, we introduce a state vector $\mathbf u = \{\mathbf w_1, \mathbf w_2, \mathbf w_3, \mathbf w_4\}$ with four separate components representing the diagonal, off-diagonal elements of the density matrix, and the position and momentum vectors of the simple harmonic oscillators, respectively. Explicitly they are defined as
\begin{eqnarray}
    & & \mathbf w_1 = \{\rho_{11}, \rho_{22}, \cdots, \rho_{LL} \}, \nonumber \\
    & & \mathbf w_2 = \{\rho_{12}, \rho_{13}, \cdots, \rho_{23}, \rho_{24}, \cdots, \rho_{L, L-1}\}, \nonumber \\
    & & \mathbf w_3 = \{Q_1, Q_2, \cdots, Q_L \}, \nonumber \\
    & & \mathbf w_4 = \{P_1, P_2, \cdots, P_L \}.
\end{eqnarray}
We partition the density matrix $\rho_{ij}$ into diagonal ($\mathbf{w}_1$) and off-diagonal ($\mathbf{w}_2$) components to highlight their conceptual distinction: the diagonal elements $\rho_\text{diag}$ represent on-site electron densities, while the off-diagonal elements $\rho_\text{off-diag}$ encode quantum coherence.

The semi-classical dynamics of the Holstein model can be cast into a form of the standard ordinary differential equation:
\begin{eqnarray}
    \label{eq:ODE_u}
	\frac{d\mathbf u}{dt} = \bm{\mathcal{F}}(\mathbf u),
\end{eqnarray}
with proper initial conditions for each component of $\mathbf{u}$. Here, we focus on time-invariant dynamical systems, as is the case for the Holstein model. By introducing a discrete prediction time step $\Delta t$, an RNN model can be trained to predict the state vector $\mathbf{u}(t + \Delta t)$ at the next step, given the current state $\mathbf{u}(t)$ as input.

The coupled differential equations in Eq.~(\ref{eq:ODE_u}) can be readily integrated to give
\begin{eqnarray}
    \mathbf u(t+\Delta t) = \mathbf u(t) + \int_t^{t+\Delta t} \boldsymbol{\mathcal{F}}(\mathbf u(t')) dt'
\end{eqnarray}
In our standard CNN approach, a single model  is trained to directly predict the second integral term in the above equation, i.e.
\begin{eqnarray}
    \label{eq:mapping1}
    \int_t^{t+\Delta t} \boldsymbol{\mathcal{F}}(\mathbf u(t')) dt' \approx \boldsymbol{\mathcal{N}}\Bigl[ \mathbf u(t) \, \Big| \, \bm \theta \Bigr]
\end{eqnarray}
where $\boldsymbol{\mathcal{N}}[\cdot , | , \bm{\theta}]$ denotes the CNN approximation of the one-step time evolution, with $\bm{\theta}$ representing its trainable parameters. The state vector at the next time-step is then approximated as
\begin{eqnarray}
	\mathbf u(t+\Delta t) \approx \mathbf u(t) + \boldsymbol{\mathcal{N}}\Bigl[ \mathbf u(t) \, \Big| \, \bm \theta \Bigr].
\end{eqnarray}

%We separate $\mathbf u$ into component parts, calculate the loss for each part, and then sum the losses. We do this to equally weigh the components. This is because using $\mathbf u$ directly in the loss would apply an implicit weight of $L$ (here denoting the system size, not the loss function) on any matrix-valued component relative to any vector-valued component, since matrix-valued components have $L^2$ elements while vector-valued ones have only $L$ elements.

%Denote the set of the separate components used for the loss calculation as $C = \{ \rho_\text{diag}, \rho_\text{off-diag}, Q, P\}$. $\rho_\text{diag}$ and $\rho_\text{off-diag}$ are the diagonal and off-diagonal elements of the density matrix $\rho$, respectively.

% For any component $c\in C$, let $c(t)$ be the ground-truth value of that component at time $t$, and let $\hat{c}(t)$ be the prediction of the \textit{update} to that component at the time-step following time $t$ obtained from the CNN. In other words, that $\hat{c}(t)$ be one of the components $\hat{\rho}_\text{diag}(t), \hat{\rho}_\text{off-diag}(t), \hat{Q}(t), \hat{P}(t)$ from the prediction $\mathcal{N}\bigl[ \mathbf u(t) \, \big| \, \bm \theta \bigr]$.
  
To train the CNN model, we define a loss function based on the four main components $\mathbf w_m$  $(m=1, \cdots 4)$. For each component, let $\delta\hat{\mathbf w}_m$ be the update or difference of the $m$-th component predicted by the CNN model~$\boldsymbol{\mathcal{N}}\bigl[ \mathbf u(t) \, \big| \, \bm \theta \bigr]$. The loss function for the CNN is then given by:
\begin{eqnarray}
	\label{eq:standard_loss_func}
	\mathcal{L}(\bm \theta \, | \, \mathbf u) =  \sum_m \sum_t \norm{ \mathbf w_m(t+\Delta t) - \mathbf w_m(t) - \delta\hat{\mathbf w}_m(t) }_2, \nonumber \\
\end{eqnarray}
where $\mathbf u(t)$ denote the ground-truth trajectory of the state vector obtained from direct numerical simulations.

\subsection{Neural network implementation} \label{sec:standard_cnn_architecture}

The standard CNN-based architecture employs a single CNN to model all three dynamical variables $Q$, $P$, and $\rho$. The network takes $\rho$, $Q$, and $P$ as input and outputs the corresponding updates $\Delta \rho$, $\Delta Q$, and $\Delta P$. A schematic diagram of the recurrent structure for the standard CNN is shown in FIG.~\ref{fig:standard_schematic}. We adopt a modified ResNet-v2~\cite{he2016identitymappingsdeepresidual} for the implementation of CNN. This choice is motivated by the residual learning framework of ResNet-v2, which facilitates the training of deep networks by mitigating vanishing-gradient issues, and by its demonstrated empirical stability and robustness across a broad range of applications. The specific details of our implementation are illustrated in FIG.~\ref{fig:standard_CNN_architecture}.

The CNN is a two-dimensional, real-valued network with four input and output channels. For a Holstein model of size $L$, the complex-valued $L \times L$ density matrix $\rho$ is first transformed into a real-valued $L \times L \times 2$ tensor, where the real and imaginary parts of $\rho$ occupy separate channels. The real-valued vectors $Q$ and $P$, each of length $L$, are embedded along the diagonals of $L \times L \times 1$ tensors, with zeros filling all off-diagonal entries. These three tensors - the $L \times L \times 2$ tensor from $\rho$ and the two $L \times L \times 1$ tensors from $Q$ and $P$ - are then concatenated along the channel dimension to form a single $L \times L \times 4$ tensor, which serves as input to the CNN.

The CNN produces an output tensor of the same size as its input, $L \times L \times 4$. The predicted increments $\Delta \rho$, $\Delta Q$, and $\Delta P$ are then extracted by reversing the embedding procedure used for the input. Specifically, $\Delta \rho$ is reconstructed by combining the first and second channels of the output tensor into a complex-valued $L \times L$ matrix, while $\Delta Q$ and $\Delta P$ are obtained from the diagonals of the third and fourth channels, respectively; see FIG.~\ref{fig:standard_CNN_architecture}. 

The internal CNN architecture is based on ResNet-v2~\cite{he2016identitymappingsdeepresidual} with several modifications.  Three modifications were made to the ResNet-v2 based internal CNN structure. First, batch normalization~\cite{ioffe2015batchnormalizationacceleratingdeep} layers were replaced with layer normalization~\cite{ba2016layernormalization} layers. Batch norm, which is commonly used in CNNs, normalizes the mean and variance of features within each mini-batch of inputs. Alongside other benefits, this helps with training stability. However, it assumes that each element of the mini-batch are sampled i.i.d from the data distribution, a condition which is not met when the training process involves multiple temporally correlated states (as is expected when training RNNs). Therefore, layer norm, which normalizes each sample independently across its features and makes no assumptions about correlations between samples of a mini-batch, is used as a drop-in replacement.

Second, we introduce channel-wise dropout~\cite{tompson14} layers preceding the final linear layer of each residual block in the CNN. These layers randomly zero a certain proportion of channels during training, and act as a regularization technique which prevents the model from depending too heavily on any particular channel. Although the original ResNet-v2 structure does not employ any form of dropout, we find its addition to be critical in improving the model's accuracy and stability. We utilize channel-wise dropout instead of traditional element-wise dropout - which works by zeroing a random proportion of \textit{all} incoming values to zero - because the random dropout of all values can introduce noise to convolutional layers by interfering with local relationships, thereby reducing overall effectiveness. Channel-wise dropout addresses this by dropping out entire channels in a convolutional layer, thereby fully preserving important locality information in the remaining channels.

Third, the Tanh (hyperbolic tangent) activation function was used instead of the original ReLU (rectified linear unit) in ResNet-v2. This was due to observed beneficial effects of Tanh to model stability and accuracy. This may be a result of Tanh being both smooth and bounded while ReLU is neither; as a result, Tanh may provide a more suitable inductive bias for regression tasks which are both smooth and empirically bounded, such as {approximating integration on the Holstein model}.

Along with the aforementioned modifications, we also employ \textit{circular} padding in our convolutional layers, as this naturally fits the periodic boundary conditions used for the simulations of the Holstein model in all state variables. Additionally, all inputs/outputs of the model contain data scaling coefficients. These are simple scalar constants set at the beginning of training based on training data statistics which normalize inputs and outputs such that the greatest absolute value in the training dataset should be $1$~\cite{Goodfellow-et-al-2016}. Normalization of inputs is common practice which can help to ensure the different components of the input have the same scale and that the range of the input is located near the region of the activation function that is most expressive. Coefficients are set separately for $\rho$, $Q$, and $P$. For example, if the largest absolute value in $\rho$ within the training dataset is found to be $0.88$, the largest absolute value in $Q$ is $1.62$, and the largest absolute value in $P$ is $0.75$, then all inputs to the CNN would have the $\rho$ component scaled by $\frac{1}{0.88}$, the $Q$ component scaled by $\frac{1}{1.62}$, and the $P$ component scaled by $\frac{1}{0.75}$. These scaling coefficients were not shown in figure~\ref{fig:standard_CNN_architecture} to reduce complexity, however, more details and a diagram can be found in appendix~\ref{sec:model_scaling_coefficients}.

\begin{figure*}[t]
\includegraphics[width=1.99\columnwidth]{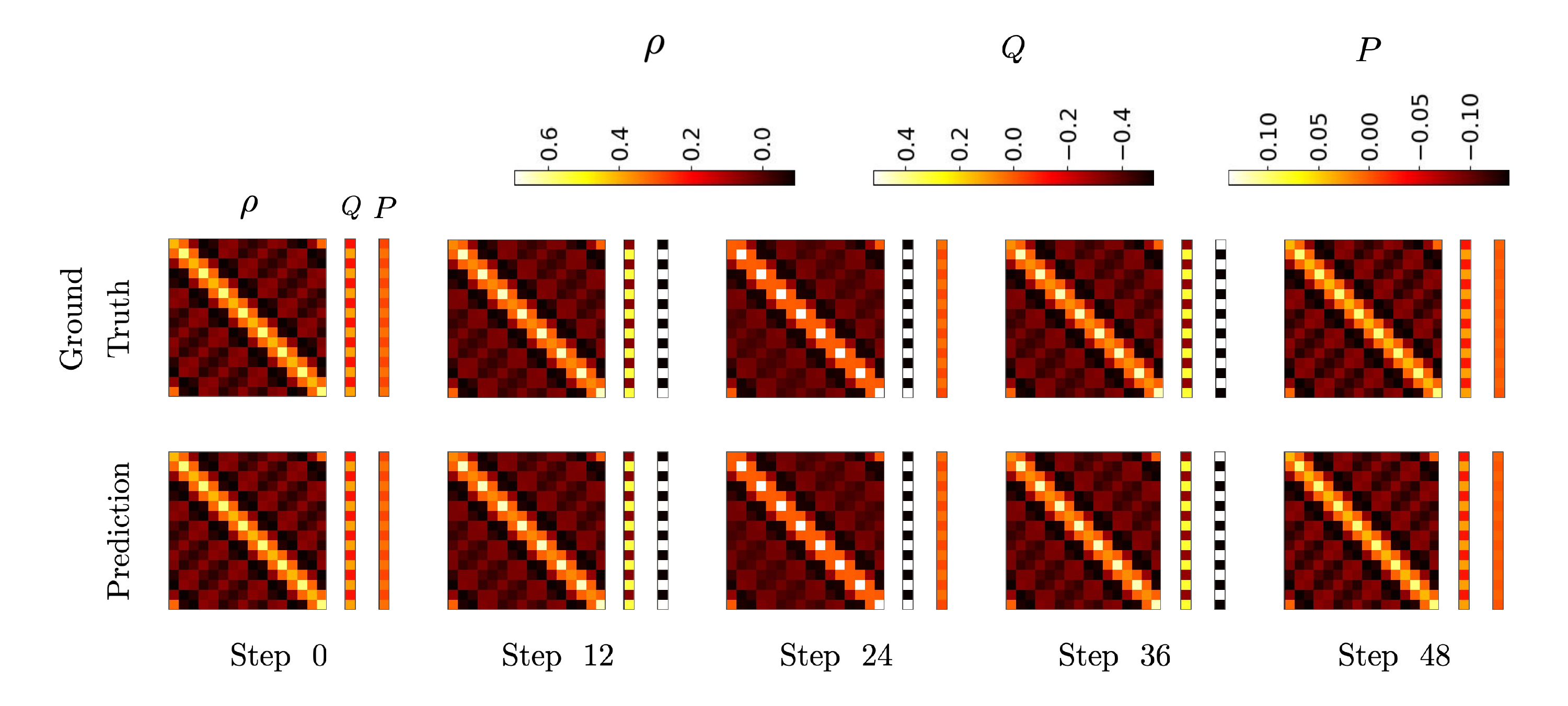}
\caption{\label{fig:shallow_visualization}
Snapshots of phonon displacements $Q_i$, momentum $P_i$, and electron density matrix $\rho_{ij}$ at various times after a shallow quench from $g_i = 0.5$ to $g_f = 0.8$ of the semi-classical Holstein model on a $L = 16$ chain. The ground truth on the top row is obtained from the 4th-order Runge-Kutta integration of the governing dynamics Eqs.~(\ref{eq:newton_eq}) and (\ref{eq:von-neumann}). The bottom row shows predictions from the trained simple CNN model. Only the real component of the complex-valued $\rho$ is shown. The steps are in units of prediction time steps, which for the shallow quench case is set to $\Delta t = 0.64$ time units. The model is given the system state at prediction step $0$ and predicts for the following $48$ prediction steps.
}
\end{figure*}

\subsection{Shallow quench data generation} \label{sec:shallow_data}

The training dataset was generated from direct numerical simulations of the semi-classical Holstein model. Specifically, the coupled Newton equation~(\ref{eq:newton_eq}) for classical phonons and the von Neumann equation~(\ref{eq:von-neumann}) for the electron density matrix were integrated using the standard fourth-order Runge-Kutta method. We pair the shallow quench scenario with the standard CNN to illustrate the relative simplicity of this regime, while the more complex deep quench scenario is modeled using the PARC-based approach, as discussed in Section~\ref{sec:PARC_CNN}.

In the shallow quench scenario, the system is initially prepared in the CDW ground state stabilized by a finite electron-phonon coupling $g_i > 0$, which is then suddenly increased at $t=0$ to a final $g_f > g_i$ with a relatively small ratio $g_f/g_i \gtrsim 1$. This abrupt change induces transient coherent oscillations of the CDW order parameter, a nonequilibrium phenomenon observed in interaction quenches of various symmetry-breaking systems, including CDW, spin-density wave, and superconducting states. Importantly, the subsequent time evolution after a shallow quench is nearly deterministic. Consequently, this dataset provides an upper bound for the model’s ability to learn simple trajectories, serving as a benchmark for assessing its generalization performance.

To generate shallow quench data, a quench from $g_i=0.5$ to $g_f=0.8$ was used. The integration time step was set to $\delta t = 0.01$ time units. Because the machine learning model can learn to predict a time step longer than the integration time step, a single prediction time step was saved every $64$ integration time steps; ie. $\Delta t = 64 \, \delta t = 0.64$. This prediction time step length was chosen to reasonably balance efficiency with effectiveness; a longer prediction interval results in less compute when using the model to approximate the integration over a given time interval, however it may become difficult for the model to learn underlying dynamics if the prediction interval is too long. We present an analysis of the effects of the prediction interval on prediction error in appendix~\ref{sec:pred_interval_scaling_analysis}. A total of $1201$ prediction time steps were saved for each trajectory, representing an initial state followed by $1200$ prediction time steps. Accordingly, this represented a domain spanning $1200\, \cdot\, 0.64 = 768$ time units. We captured $64$ trajectories in total, each offset by one integration time step to provide uniform temporal sampling throughout the prediction interval of $\Delta t=0.64$.

\subsection{Training details} \label{sec:standard_training_implementation}

Given the dataset in the form of exact system trajectories $\mathbf u(t) = [ \rho(t), Q(t), P(t) ]$, the loss function defined in Eq.~(\ref{eq:standard_loss_func}) is used to obtain the optimized parameters $\bm \theta^*$ in the standard CNN model case. As can be seen in the loss function, the overall loss is the sum over \textit{multiple} time steps. This is because we use multi-step prediction while training, wherein for some step value $N \geq 1$, the model recurrently generates predictions for prediction time steps $t+\Delta t, t+2\Delta t, \dots, t+N\Delta t$ when given the input at prediction time step $t$. The resulting predictions for the $N$ prediction time steps are then compared to the label values for the corresponding $N$ prediction time steps in the data for the calculation of the loss. We find that employing multi-step prediction greatly improves the model's accuracy, likely because it forces the model to adapt to its own errors and better learn the long-term dynamics in the data.

The AdamW optimizer~\cite{loshchilov17}, which is a variant of the popular Adaptive Moment Estimation (Adam) algorithm~\cite{kingma17}, is used for gradient descent minimization of the loss function. AdamW is a simple modification of Adam which handles the weight decay - a technique to preventing overfitting by penalizing large weights in neural nets - correctly. Weight decay helps improve model generalization as large weight values can often indicate a model is "relying" on specific, memorized features within training data instead of properly generalizing.

Alongside AdamW and weight decay, a few additional methods are used to stabilize and improve training. We use gradient clipping, a common technique which clips the gradient calculated during training to a maximum norm threshold to prevent massive and destabilizing updates to the model weights. This is an especially notable concern while training RNNs, as training with multi-step prediction can result in exploding gradients (gradients which accumulate to enormous values)~\cite{pascanu2013difficultytrainingrecurrentneural}.

During training, we add a small amount of gaussian noise to the input of the model. The motivation behind this is to improve the model's ability to adapt to and correct deviations from the training data distribution, which is unavoidable as predictions errors compound during long recurrent prediction sequences. We find that the addition of this noise during training is highly beneficial to the model's generalization abilities.

We employ curriculum learning~\cite{bengio2009curriculum} while training, which gradually increases the difficulty of the training process as the model learns. Specifically, we gradually increase the magnitude of the gaussian noise added to the input and the number of steps used in multi-step prediction with each stage of the curriculum. To compliment curriculum learning, we use a learning rate scheduler (which modifies the learning rate during training) which implements cosine annealing with warm restarts~\cite{loshchilov2017sgdrstochasticgradientdescent}, and we align the restarts with the stages of our curriculum. Essentially, this learning rate scheduler maximizes the learning rate at the start of each stage in the curriculum and then slowly decreases the learning rate following a cosine curve, allowing the model to quickly learn at the start and then slow down updates as its weights converge. In addition, we further augment the learning rate scheduler by using a short linear learning rate warm-up at the beginning of each restart, which warms up the learning rate to its maximum value over a certain number of steps instead of at once. This is beneficial because Adam-family optimizers rely on estimates of the first and second moments of the gradient, and utilizing a warm-up reduces the updates the optimizer can make until it has learned better estimates for these values

\label{sec:shallow_results}

\begin{figure}[t]
\includegraphics[width=0.99\columnwidth]{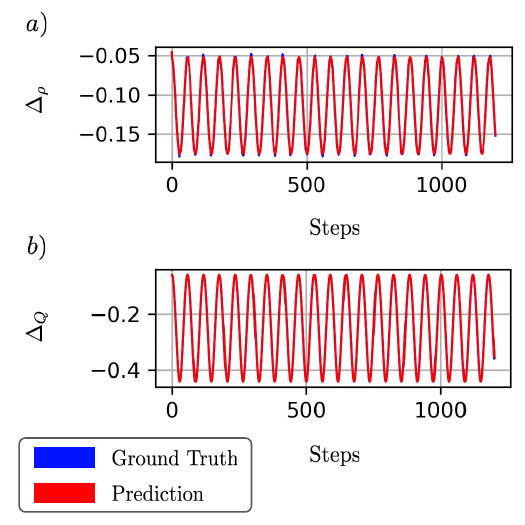}
\caption{\label{fig:shallow_cdw}
Graphed ground truth (blue) vs predicted (red) $\Delta_\rho$ and $\Delta_Q$ for a shallow quench trajectory with system size $32$. The red prediction essentially perfectly covers the blue ground truth, which is under the red prediction in the visualization.
}
\end{figure}

\subsection{Benchmark}

FIG.~\ref{fig:shallow_visualization} illustrates both the ground-truth trajectories and the predictions generated by a standard CNN model trained on shallow-quench data for a system of size $16$. Owing to the highly deterministic and nearly periodic nature of the shallow-quench dynamics, we find that even a very compact CNN model - on the order of $\sim 5$K trainable parameters - suffices to fully capture the system’s evolution.

To present a more quantitative comparison, we first introduce the time-dependent order parameters of the system. First, the CDW order, or electron-number modulation, with a wave-vector $Q = \pi /a$ can be described by
\begin{eqnarray}
    \label{eq:cdw-order}
    \Delta_{\rho}(t) = \frac{1}{L} \sum_{i} \rho_{ii}(t) \cos\left(\frac{\pi}{a} x_i\right).
\end{eqnarray}
The electron number expectation value here is directly given by the diagonal elements of the density matrix: $\langle \hat{n}_i \rangle = \rho_{ii}(t)$. The ideal CDW order described by Eq.~(\ref{eq:n-mod}) corresponds to an order parameter of $\Delta = \delta n$. The staggered lattice distortion accompanying the charge modulation can be described by a similar order parameter
\begin{eqnarray}
    \Delta_Q(t) = \frac{1}{L} \sum_i Q_i(t) \cos\left(\frac{\pi}{a} x_i\right). 
\end{eqnarray}
The time evolution of the two order parameters is shown in FIG.~\ref{fig:shallow_cdw}, comparing exact numerical simulations (ground truth) with CNN predictions. The Holstein system is initialized with the exact state and evolved forward for $1200$ prediction time steps. The CNN reproduces the periodic oscillations with essentially perfect fidelity, showing no visible deviation from the ground truth.

This striking accuracy reflects the relative simplicity of the shallow-quench regime: the deterministic dynamics can be effectively ``memorized" by a small, standard CNN. In this case, the learning task requires neither elaborate architecture nor large model capacity, underscoring that the shallow quench serves primarily as a baseline demonstration rather than a demanding test of predictive power.

\begin{figure*}[t]
\includegraphics[width=1.99\columnwidth]{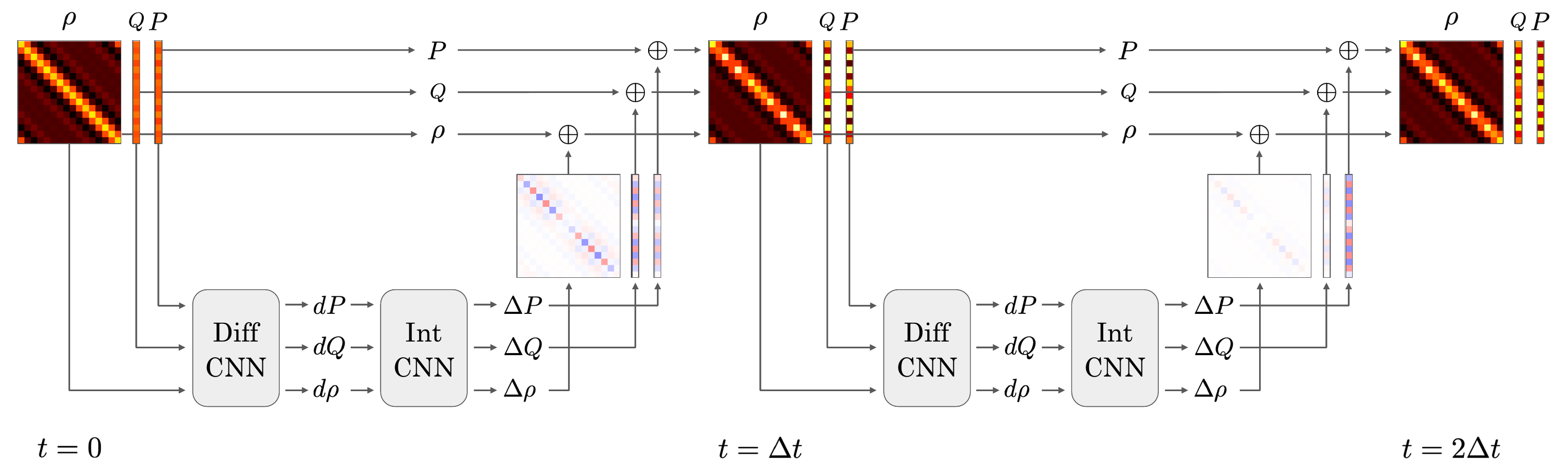}
\caption{\label{fig:parc_schematic}
Schematic diagram of the recurrent structure for non-adiabatic dynamics of the semi-classical Holstein model. The framework is based on a single PARC-based CNN differentiator (diff)/integrator (int) pair for all three dynamical degrees of freedom: $Q$, $P$, and $\rho$.
}
\end{figure*}

\section{PARC CNN and Deep Quench} \label{sec:PARC_CNN}

Although a recurrent structure based on a single standard CNN suffices to reproduce the relatively deterministic dynamics of the Holstein model following a shallow quench, capturing more complex chaotic spatiotemporal evolution motivates the incorporation of additional physical constraints. To address this, we consider the deep quench protocol, wherein initially decoupled lattice and electronic subsystems (with $g_i = 0$) are abruptly coupled through a large nonzero interaction $g_f \neq 0$. The ensuing dynamics are characterized by pronounced chaotic behavior. Owing to the accumulation of prediction errors, it is inherently infeasible for machine learning models to resolve long-time trajectories of such chaotic systems. Nonetheless, we demonstrate that a recently proposed Physics Aware Recurrent Convolution (PARC) architecture is capable of faithfully reproducing the statistical properties, or “climate,” of the spatiotemporal dynamics generated by the deep quench.

\subsection{PARC architecture}

We begin by discussing the PARC neural network architecture, originally introduced to model the dynamics of highly nonlinear PDEs~\cite{nguyen23}.  Rather than employing a single neural network to approximate the mapping $\mathbf u(t) \mapsto \mathbf u(t + \Delta t)$, PARC introduces two convolutional neural networks, which play the roles of a differentiator and an integrator, respectively, in analogy with the two stages of a finite-difference scheme for solving differential equations. This design is directly motivated by the numerical integration of differential equations over a finite time interval $\Delta t$.
\begin{eqnarray}
	\mathbf u(t + \Delta t) = \mathbf u(t) + \int_{t}^{t+\Delta t} \frac{d \mathbf u}{dt} dt,
\end{eqnarray}
where the time derivative $d\mathbf u/dt$ is governed by the $\boldsymbol{\mathcal{F}}(\mathbf u)$ function defined in Eq.~(\ref{eq:ODE_u}). 

In the shallow quench case discussed in Sec.~\ref{sec:standard_CNN}, a single CNN is introduced to approximate the second integral term in the above equation. The motivation of the PARC approach is to incorporate the two-step operations: differentiation followed by integration, into the recurrent model. 
To this end, we first introduce a CNN model $\boldsymbol{\mathcal{D}}[\cdot|\bm \phi]$ to approximate the differential operation
\begin{eqnarray}
    \boldsymbol{\mathcal{D}}[\mathbf u|\bm\phi] \approx \frac{d\mathbf u}{dt},
\end{eqnarray}
where $\bm \phi$ are the trainable parameters of the CNN model. In our benchmark study, the time derivative is explicitly provided by the function $\boldsymbol{\mathcal{F}}(\cdot)$. However, introducing the differentiator CNN enables us to extend the RNN framework to data-based empirical dynamics where no explicit model is available. Next we introduce another CNN-based model for the integration operation:
\begin{eqnarray}
    \boldsymbol{\mathcal{S}}[\bm f|\bm \theta] \approx \int_t^{t+\Delta t} \bm f(t') dt',
\end{eqnarray}
where $\bm\theta$ are the corresponding trainable parameters. 
The increment of the state vector over $\Delta t$ can then be expressed as the composition of differential and integral operators. Explicitly, the mapping over one time-step is given by
\begin{eqnarray}
	\mathbf u(t+\Delta t) = \mathbf u(t) 
    + \boldsymbol{\mathcal{S}}\Bigl[ \boldsymbol{\mathcal{D}}\bigl[ \mathbf u(t) \, \big| \,\bm\phi \bigr] \, \Big| \, \bm \theta \Bigr].
\end{eqnarray}
Compared to the single-CNN recurrent structure in Eq.~(\ref{eq:mapping1}), this approach decomposes the single mapping~$\boldsymbol{\mathcal{N}}$ into two distinct CNN models, corresponding to the differentiator and integrator operators, respectively.

Based on this physics motivated architecture, the PARC loss function extends the standard CNN loss Eq.~(\ref{eq:standard_loss_func}) by adding a differentiator loss component. Let ${d \hat{\mathbf w}_m}/{dt}$ represent the mid-point (between the input time step and the prediction time step) derivative of the $m$-th component predicted by the CNN differentiator $\boldsymbol{\mathcal{D}}[ \mathbf u(t) \, | \, \bm\phi ]$, and let $\delta\hat{\mathbf w}_m$ be the update or difference of the $m$-th component predicted by the CNN integrator $\boldsymbol{\mathcal{S}}\bigl[ \boldsymbol{\mathcal{D}}[ \mathbf u(t) \, |  \, \bm\phi ] \, \big| \, \bm\theta \bigr]$. The PARC loss function becomes:
\begin{eqnarray}
	\label{eq:parc_loss_func}
	& & \mathcal{L}(\bm \phi, \bm \theta \, | \, \mathbf u) 
    =  \sum_m \Biggl( \sum_t \norm{ \frac{d\mathbf w_m}{dt}\bigg|_{t+\Delta t/2} - \frac{d\hat{\mathbf w}_m}{dt}\bigg|_{t+\Delta t/2} }_2 \nonumber \\
	& & \qquad \qquad + \sum_t \norm{\mathbf w_m(t+\Delta t) - \mathbf w(t) - \delta\hat{\mathbf w}_m(t) }_2 \Biggr), \nonumber
\end{eqnarray}
where $\mathbf u(t) = \{\mathbf w_m(t)\}$ denotes the ground-truth trajectory of the state vector obtained from direct numerical simulations.

\begin{figure*}[t]
\includegraphics[width=1.99\columnwidth]{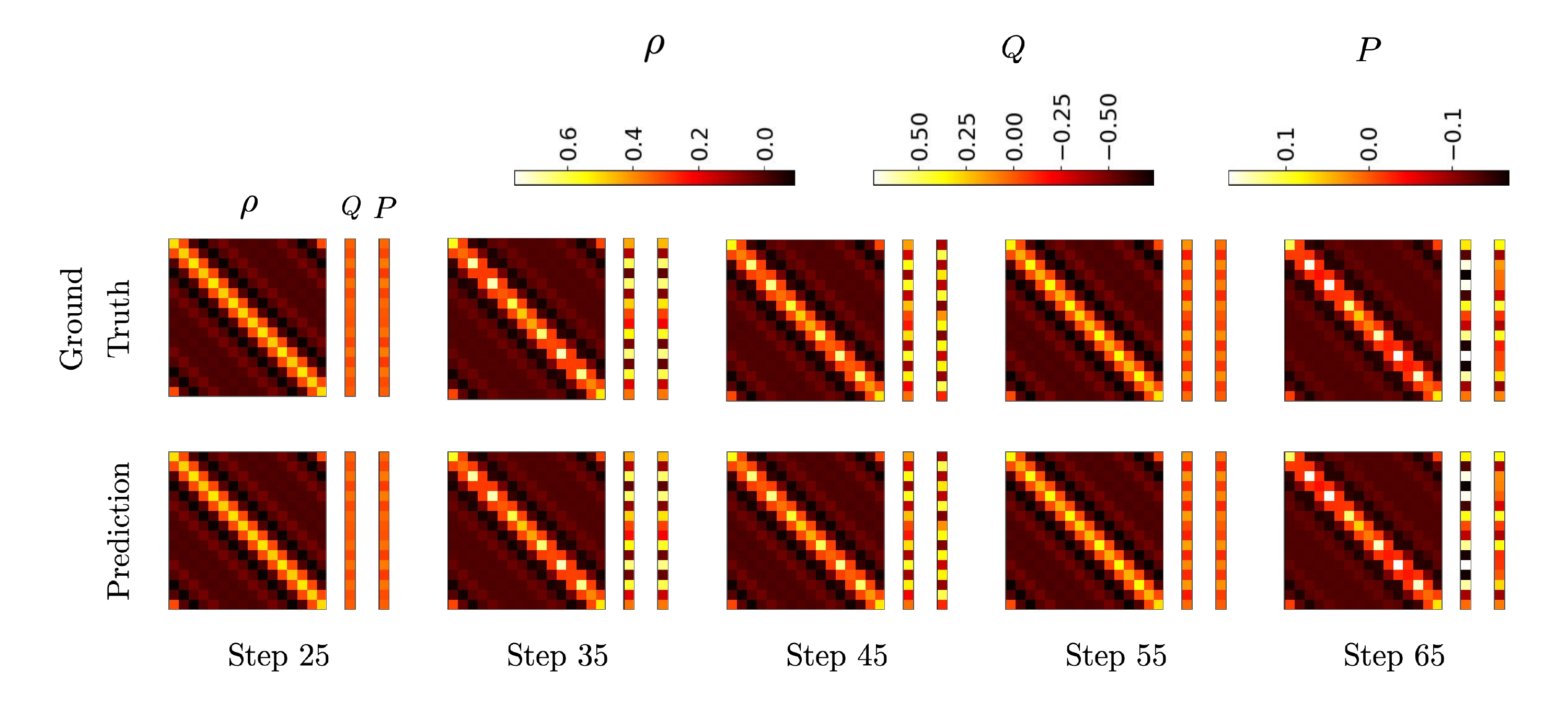}
\caption{\label{fig:deep_visualization}
Snapshots of phonon displacements $Q_i$, momentum $P_i$, and electron density matrix $\rho_{ij}$ at various times after a deep quench from $g_i = 0.0$ to $g_f = 1.0$ of the semi-classical Holstein model on a $L = 16$ chain. The ground truth on the top row is obtained from the 4th-order Runge-Kutta integration of the governing dynamics Eqs.~(\ref{eq:newton_eq}) and (\ref{eq:von-neumann}). The bottom row shows predictions from the trained PARC-based model. Only the real component of the complex-valued $\rho$ is shown. The steps are in units of prediction time steps, which for the deep quench case is set to $\Delta t = 2.56$ time units. Step $25$ correlates to $64$ time units after the quench, which is the point by which the system enters its post-transient regime. The model predictions start at prediction step $25$ and continue for $40$ predictions steps until step $65$.
}
\end{figure*}

\subsection{Architecture Implementation Details} \label{sec:parc_model_architecture}

We next discuss details of the PARC implementation. A schematic diagram of the recurrent structure of our PARC-based model is shown in FIG.~\ref{fig:parc_schematic}. As described earlier, PARC introduces two CNNs corresponding to the differentiator and integrator steps of a finite-difference method for solving differential equations.

A single PARC-based neural network is constructed to evolve all three dynamical variables, $\rho$, $Q$, and $P$. Both the differentiator and integrator components are structurally identical CNNs, sharing the same architecture as the standard CNN described in Sec.~\ref{sec:standard_cnn_architecture}. The input is prepared as follows: the complex-valued $L\times L$ matrix $\rho$ is decomposed into a real-valued $L\times L\times 2$ tensor, while $Q$ and $P$ are placed along the diagonals of $L\times L\times 1$ tensors. Concatenating these yields an $L\times L\times 4$ input tensor.

Each CNN (differentiator or integrator) maps this input to an output tensor of the same dimension. The physical quantities of interest - $\frac{d\rho}{dt}$, $\frac{dQ}{dt}$, $\frac{dP}{dt}$ in the case of the differentiator, and $\Delta\rho$, $\Delta Q$, $\Delta P$ in the case of the integrator - are then obtained by reversing the embedding procedure: recombining the first two channels into a complex $L\times L$ matrix for $\rho$ and extracting $Q$ and $P$ from the diagonals of the remaining channels. A schematic of this data flow is shown in Fig.~\ref{fig:standard_CNN_architecture}. Additionally, we calculate data scaling coefficients from the data and integrate them into the model's inputs and outputs as a form of data normalization. Although they are part of the model, we did not include them in Fig.~\ref{fig:standard_CNN_architecture} to reduce complexity. Further details regarding the data scaling coefficients can be found in appendix~\ref{sec:model_scaling_coefficients}.

\subsection{Deep Quench Data} \label{sec:deep_data}

As with the shallow quench data discussed in section~\ref{sec:shallow_data}, the deep quench training dataset was obtained from direct numerical simulations of the semi-classical Holstein model. In the deep quench scenario, the coupling is initially turned off $g_i = 0$, which means the corresponding ground state is free electron gas on a chain decoupled from a set of independent simple harmonic oscillators in equilibrium $Q_i = P_i = 0$. Naturally, there is no CDW order in this initial state.  The electron-phonon coupling is then suddenly switch on to $g_f > 0$ at time $t = 0$. The quench dynamics in this scenario is dominated by the emergence of CDW orders induced by the nonzero $g_f$. The onset of the CDW order is a stochastic process. First local CDW order is initiated through nucleation process at random seeds. This is then followed by the growth and merger of CDW domains. This scenario is akin to the phase-ordering process when a system is suddenly quenched from a high-temperature random state into a symmetry-breaking phase. As a result, the deep-quench process is very sensitive to random fluctuations in the initial states. For a controlled approach to generate dataset, $Q_i$ are sampled from a normal distribution of zero mean and a standard deviation of $10^{-4}$.  Contrary to the shallow quench scenario discussed above, a wide variety of dataset with different system trajectories can be generated from the deep quench simulations.

To generate the deep-quench dataset, we performed a quench from $g_i = 0.0$ to $g_f = 1.0$, yielding a total of $1228$ trajectories. While the integration time step used to generate the data is still $\delta t = 0.01$ time units, we set the prediction time step to $\Delta t = 256\, \delta t = 2.56$ time units. For the deep quench, snapshot capture began only after the first $64$ time units, by which point the system had reached its post-transient regime. After this, we capture a total of $1001$ prediction time steps (an initial state and $1000$ steps), totaling a time domain per trajectory of $1000\,\cdot\,2.56 = 2560$ time units. To enable the calculation of derivatives between prediction time steps - as seen in the differentiator term of the PARC loss function as shown in Eq.~(\ref{eq:parc_loss_func}) -, we additionally record the intermediate states between successive prediction time steps. Finally, in a similar fashion to the generation of the shallow quench data, we offset the starting point of each trajectory by a certain number of integration time steps to provide uniform temporal sampling throughout the prediction time interval of $\Delta t = 2.56$.

\subsection{Training Details} \label{sec:parc_training_implementation}

The training procedure for the PARC-based model is a direct extension of that for the standard CNN model described in Sec.~\ref{sec:standard_training_implementation}, with only minor modifications to accommodate the additional structure of PARC.

Given datasets in the form of exact system trajectories $\mathbf u(t) = [\rho(t), Q(t), P(t)]$, the PARC loss function, Eq.~(\ref{eq:parc_loss_func}), is minimized via gradient descent to obtain the optimized parameters $\bm \phi^*$ and $\bm \theta^*$. Training proceeds with multi-step prediction: for each trajectory segment, $N$-step predictions are compared against the corresponding ground-truth states, while the differentiator simultaneously produces $N$ intermediate derivative estimates. These are evaluated against derivatives computed from the mid-interval states, providing the additional supervision required by the PARC framework.

\begin{figure}[t]
\includegraphics[width=0.99\columnwidth]{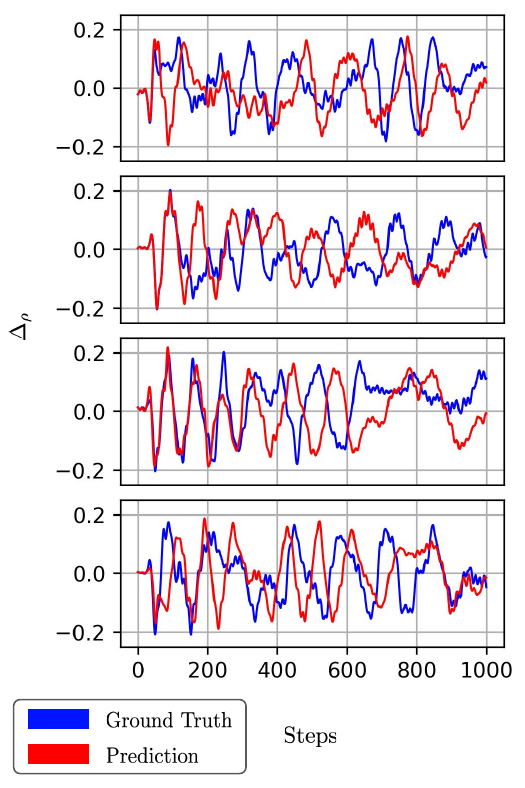}
\caption{\label{fig:deep_quench_cdw_traces}
Comparison of $\Delta_\rho$ traces between ground truth data and prediction trajectories starting from the same initial condition. The ground truth trajectories/initial conditions are randomly selected from a deep quench dataset with system size $32$. Blue traces are from ground truth data and red traces are from model predictions. Model predictions were made for $1000$ steps after the model was provided with a ground truth starting state. Steps are in units of $\Delta t = 2.56$ time units. The model predictions adhere well to the ground truth traces in the short-term, but diverge in the long-term.
}
\end{figure}

\section{Deep Quench Results} \label{sec:deep_results}

FIG.~\ref{fig:deep_visualization} visualizes a ground-truth trajectory and predictions from a PARC-based model trained on a deep quench dataset with system size $L=16$. The model was given the state at prediction step $25$ as input (which corresponds to the start of the saved deep quench dataset following the initial 64 time units of numerical integration during the system's transient regime) and then predicted for $40$ steps thereafter to create the visualization. As previously noted, the deep quench trajectories are highly sensitive to initial condition, and small perturbations to the initial condition result in different system trajectories. All trajectories used in the analysis of the deep quench case belonged to a test set not shared with the training set. Additionally, both ground truth trajectories and predictions have initial states starting after the transient regime, which, as noted in Sec.~\ref{sec:deep_data}, begins $64$ time units after the quench.

The shared initial state in both cases is characterized by nearly constant diagonal elements $\rho_{ii} \sim \text{const.}$, indicating a uniform charge distribution, together with approximately uniform lattice displacements $Q$ and momenta $P$. As discussed above, this configuration corresponds to decoupled electron and lattice subsystems close to their respective ground states. Once the coupling is switched on at $g_f = 1$, energetic considerations suggest that the system tends to develop CDW order. Because the formation of CDW order in distant regions is largely independent, owing to the local nature of the symmetry-breaking process, the system is expected to evolve into an inhomogeneous state consisting of multiple CDW domains of opposite sign, separated by domain walls. Indeed, while the electron density remains relatively uniform, as indicated by the diagonal line of the density matrix, a pronounced inhomogeneous CDW state emerges at later times following the quench (see Fig.~\ref{fig:deep_visualization}), becoming visible by time-step $25\Delta t$. 

Remarkably, even for such strongly inhomogeneous space-time dynamics, the ML predictions remain in close agreement with the ground truth at $40\,\Delta t$ after the beginning the prediction. This highlights the ability of the model to capture highly nontrivial dynamical features well beyond the initial stages of evolution. Nevertheless, as with any statistical learning approach, prediction errors are unavoidable. While such errors may remain small at short and intermediate times, their gradual accumulation is inevitable and eventually compromises the reliability of long-time forecasts. This limitation is clearly illustrated in Fig.~\ref{fig:deep_quench_cdw_traces}, which compares the time traces of the CDW order parameter $\Delta_\rho(t)$ from four independent exact simulations with those obtained from the PARC model. Importantly, this issue is not specific to the present implementation but reflects a fundamental challenge of ML-based dynamical modeling, where small stepwise inaccuracies compound over time and obscure the true asymptotic behavior of the system.

Long-term prediction is particularly difficult in chaotic dynamical systems due to their extreme sensitivity to small variations in initial conditions. The post-quench dynamics of the semi-classical Holstein model exhibits such chaotic behavior, as shown in Fig.~\ref{fig:deep_quench_cdw_traces}: four independent $g_i = 0 \to g_f = 1$ quench simulations initiated with slightly different stochastic noise profiles produce markedly distinct trajectories of $\Delta_\rho(t)$ (blue curves). Beyond this hypersensitivity to initial conditions, the CDW order parameter exhibits no emergent structure or reproducible dynamical pattern. As a result, even a highly optimized model will begin to diverge from a ground truth trajectory over many steps.

%Figure~\ref{fig:deep_quench_cdw_traces} provides a visualization of some random CDW traces from both ground truth and prediction to provide a sense of the "random" nature of the deep quench trajectories.

\begin{figure}[t]
\includegraphics[width=0.99\columnwidth]{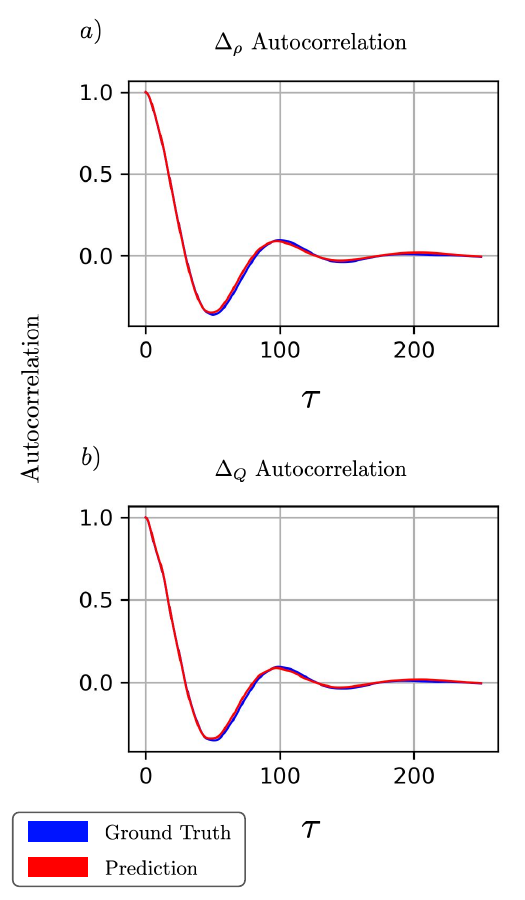}
\caption{\label{fig:deep_autocorrelation}
Ground truth vs predicted autocorrelation trajectories of both $\Delta_\rho$ and $\Delta_Q$. Values of $\tau$ up to a maximum of $250$ were used out of $1000$ total steps in the trajectories. The same starting states from the ground truth trajectories were used for the predicted trajectories. There were $256$ each of ground truth and predicted trajectories. The system size is~$32$.
}
\end{figure}

%\hlc{yellow}{Due to the highly random and dynamic nature of deep quench trajectories}, even a highly optimized model will begin to diverge from a ground truth trajectory over many steps. Because of this, we use the autocorrelation function as a better indicator of whether the model has learned the underlying physical dynamics of the data. We define the autocorrelation function as

On the other hand, while the ML predictions drift away from the ground-truth trajectories, the predicted time traces seem to retain the same statistical characteristics, implying that our ML models could successfully capture the underlying chaotic attractor dynamics. To demonstrate this, we consider the autocorrelation function for the order parameters ($X = \Delta_\rho$ or $\Delta_Q$):
\begin{eqnarray}
	\label{eq:autocorrelation}
	A(\tau) = \frac{\langle X(\tau_0)X(\tau_0 + \tau)\rangle - \langle X(\tau_0)\rangle}{\langle X(\tau_0)^2\rangle - \langle X(\tau_0)\rangle}^2 , 
\end{eqnarray}	

\begin{figure*}[t]
\includegraphics[width=1.9\columnwidth]{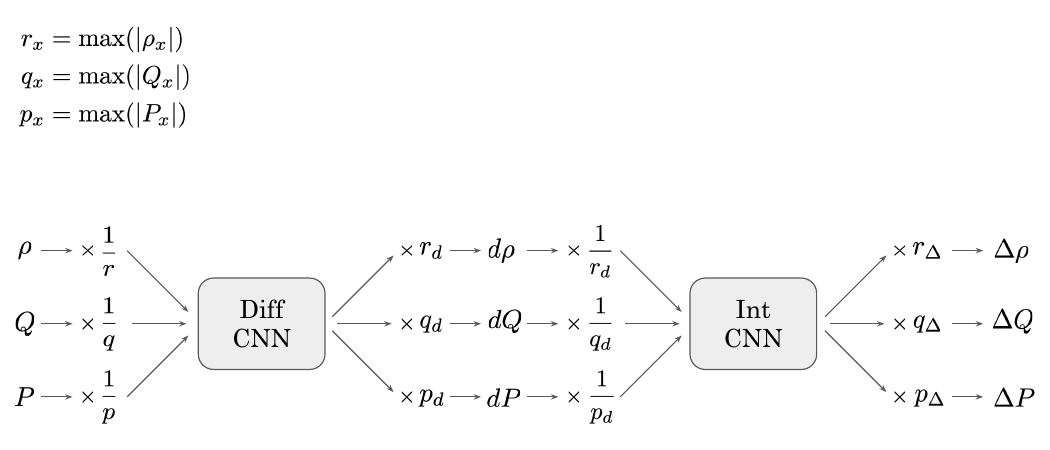}
\caption{\label{fig:data_scalars}
An illustration showing the application of data scaling coefficients on the input and output of the model.
}
\end{figure*}
where $\tau$ is the time lag, and the brackets $\langle \cdots \rangle$ denote averaging over both the initial time $t_0$ and an ensemble of independent runs, which together approximate an average over the invariant measure of the time series. The autocorrelation function provides a quantitative measure of temporal correlations in a dynamical trajectory and is particularly useful for diagnosing chaotic behavior. By definition, $A(0) = 1$, since a trajectory is perfectly correlated with itself at zero lag. As $\tau$ increases, $A(\tau)$ decays, reflecting the gradual loss of temporal memory. For trajectories evolving on a chaotic attractor, $A(\tau)$ typically approaches zero at large lag times, consistent with the system’s sensitivity to initial conditions and its ergodic exploration of the attractor. In this sense, the autocorrelation function probes not only local dynamical behavior but also the invariant measure that governs long-time statistics.

We apply this analysis to the charge-density-wave order parameter $\Delta_\rho(t)$ and the lattice displacement $\Delta_Q(t)$, comparing ground-truth trajectories with those generated by the ML model. For each case, $256$ independent trajectories of length $1000$ time steps were analyzed, with lag values considered up to $250$ steps. The resulting autocorrelation curves are shown in FIG.~\ref{fig:deep_autocorrelation}. Strikingly, the predicted trajectories reproduce the autocorrelation functions of the ground truth with high fidelity. This agreement demonstrates that the ML model has not only learned to reproduce individual short-time trajectories, but also captured the ergodic properties and invariant statistical structure of the underlying chaotic dynamics.

%where $X(\tau_0)$ is the entirety of a 1D trajectory/sequence starting at time $\tau_0$, $X(\tau_0 + \tau)$ is the portion of the trajectory starting $\tau$ time after the start at $\tau_0$, $\langle \cdot \rangle$ is an average over both samples of a batch as well as time, and $X(\tau_0)X(\tau_0+\tau)$ represents the dot product between the portion of $X$ starting at $\tau_0$ and ending $\tau$ time before the end of the trajectory, and the portion of $X$ defined by $X(\tau_0 + \tau)$. The time $\tau$ mentioned here is referred to as the "lag" value, and is in units of prediction time steps.

%Conceptually, this autocorrelation function is a metric which defines how correlated a trajectory is with itself when separated by a certain period of lag. When the lag is $0$, the autocorrelation is $1$. \hlc{yellow}{As lag $\to \infty$, a trajectory with dynamics which shift over time should see the autocorrelation $\to 0$.} We apply and compare the autocorrelation function to the $\Delta_\rho$ and $\Delta_Q$ trajectories of both ground truth data and model predictions. $256$ trajectories are used for each the ground truth and prediction. Each trajectory contains $1000$ steps, and the maximum lag value is $250$. As is shown in figure~\ref{fig:deep_autocorrelation}, the resulting autocorrelation curve of the predicted trajectories closely match that of the ground truth, indicating that the model has learned excellent physical dynamics from the training data.

\begin{figure*}[t]
\includegraphics[width=1.9\columnwidth]{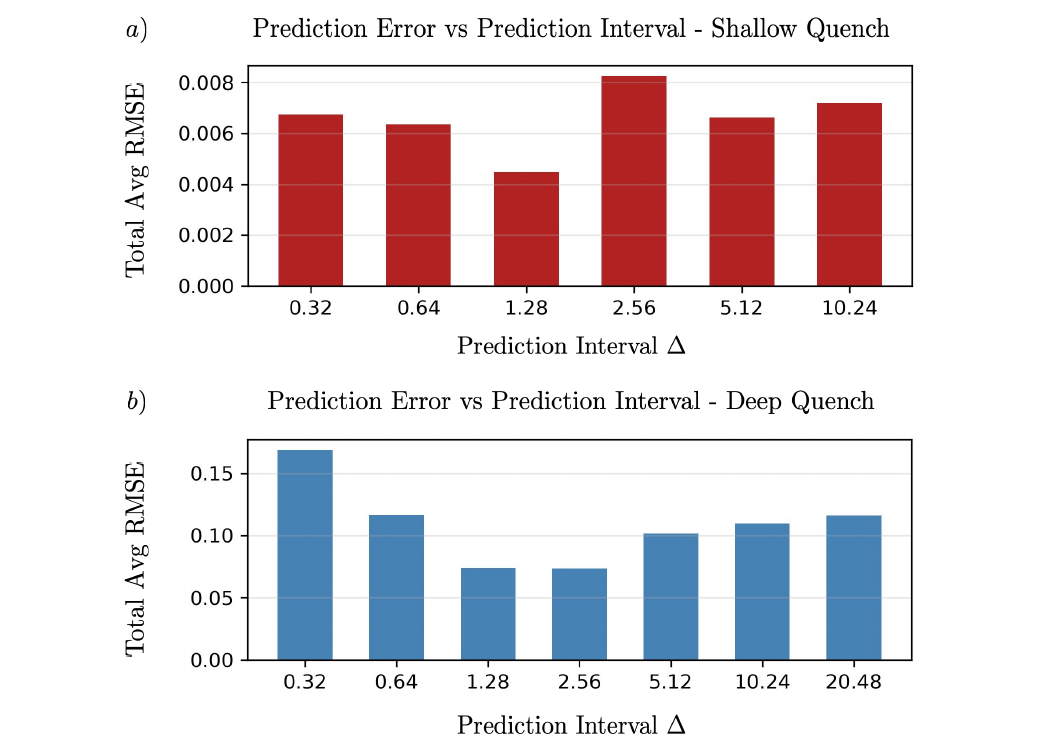}
\caption{The prediction error of different models over a time interval of $81.92$ time units. Different models were trained with different prediction intervals ranging from $\Delta = 0.32$ to $\Delta = 10.24$. Models with shorter prediction intervals required more steps to predict over the same interval of time. The "Total Avg RMSE" represents the sum of the average RMSE for each of the components $\rho$, $Q$, and $P$ over $256K$ random prediction targets from a test set.}
\label{fig:scaling_analysis}
\end{figure*}

\section{Summary and Outlook} \label{sec:summary}

We have demonstrated that recurrent convolutional architectures can effectively model the nonequilibrium dynamics of the one-dimensional semi-classical Holstein model across distinct quench protocols. In the case of the shallow quench, where the dynamics are largely deterministic and exhibit simple periodic structure, we showed that even an extremely compact recurrent CNN, with only a few thousand trainable parameters, is sufficient to accurately reproduce the system’s time evolution. This highlights the efficiency of standard deep-learning architectures in capturing relatively simple dynamical regimes.

By contrast, the deep quench scenario presents a far greater challenge, as the sudden onset of strong electron-phonon coupling drives chaotic spatiotemporal dynamics that defy accurate trajectory-level prediction. For this regime, we introduced a Physics Aware Recurrent Convolutional (PARC) architecture, trained solely on sequential state data, and demonstrated its ability to faithfully reproduce the long-term statistical ``climate” of the dynamics. The efficacy of the PARC model derives from its differentiator-integrator structure, which encodes a physics-aware inductive bias directly into the network. This hybrid design augments the CNN’s capacity to learn local hierarchical correlations with a structural prior aligned to the underlying equations of motion.

Our results show that the PARC model not only generalizes to unseen initial conditions but also achieves stable multi-step integration through recurrent prediction, allowing it to propagate solutions over long horizons from arbitrary starting points. Visualization of the predicted trajectories confirms close agreement with ground-truth simulations, both at the level of individual time evolutions and in terms of ensemble statistical properties. These findings establish that the PARC-based approach is capable of learning and reproducing physically consistent dynamics, going beyond surface-level trajectory matching to capture the underlying physical laws with quantifiable accuracy.

An extension of this work could be the scalability of the model to different system sizes. Although the model presented in this work is pure-convolution, and thus structurally scalable to any system size, in reality effective scaling is non-trivial as in the finite system size regime, local dynamics of the one-dimensional semi-classical Holstein model can vary heavily with system size, especially towards the smaller end. However, due to the quadratic scaling nature of compute with system size in CNNs, training CNN models on larger system sizes can be highly resource intensive to the point of impracticality. Therefore, either the compute requirement of training on larger system sizes must be reduced, or transfer learning between smaller and larger system sizes can be explored.

Our present treatment of non-adiabatic dynamics in hybrid quantum-classical systems follows the Ehrenfest mean-field approach~\cite{marx09,li05x}. While conceptually simple and efficient, this method neglects key features such as wave-packet branching and stochastic surface transitions. Extending machine-learning (ML) models to more sophisticated schemes~\cite{horsfield04,tapavicza13}, such as Tully's fewest-switches surface-hopping algorithm~\cite{jain22}, presents additional challenges. The PARC framework, or more generally the RNN models used here and in related studies, are designed to learn deterministic trajectories governed by coupled ordinary differential equations (ODEs). As such, they cannot be directly applied to inherently stochastic surface-hopping dynamics. Nevertheless, PARC models can be trained to capture the deterministic evolution of both classical and electronic degrees of freedom between hopping events. Achieving a fully ML-based description of non-adiabatic dynamics would require additional components capable of predicting hopping probabilities and modeling stochastic transitions of the active surface. Alternatively, one may develop RNN or PARC models to learn ensemble-averaged trajectories derived from surface-hopping simulations. These extensions represent promising directions for future research.

\begin{acknowledgments}
This work was supported by the U.S. Department of Energy, Office of Basic Energy Sciences, under Contract No. DE-SC0020330. A. P. Ning acknowledges support from the Owens Family Foundation, and L. Yang acknowledges support from the Jefferson Fellowship. The authors also acknowledge the support provided by Research Computing at the University of Virginia.
\end{acknowledgments}

%\section{Appendix}

\appendix

\section{Model Input/Output Scaling Coefficients}
\label{sec:model_scaling_coefficients}

Data normalization is commonly performed as a preprocessing step, applied to the dataset prior to being used as input to the model. This approach is often sufficient in classification tasks, where only relative differences between output values are important.

In contrast, regression tasks such as ours require the model outputs to reproduce \textit{exact} values. For this reason, it is more natural to implement normalization directly within the model architecture, rather than as part of the external training or evaluation pipeline. This design choice is particularly advantageous when training multiple models across diverse datasets, each of which may require distinct scaling factors to achieve the desired normalized range.

For every dataset, we compute nine different coefficients, each representing the maximum absolute value of some component within the dataset:
\begin{itemize}
    \item $r$: The maximum absolute value of all $\rho$
    \item $q$: The maximum absolute value of all $Q$
    \item $p$: The maximum absolute value of all $P$
    \item $r_d$: The maximum absolute value of all $d\rho/dt$
    \item $q_d$: The maximum absolute value of all $dQ/dt$
    \item $p_d$: The maximum absolute value of all $dP/dt$
    \item $r_\Delta$: The maximum absolute value of all $\Delta \rho$ (step-wise update to $\rho$)
    \item $q_\Delta$: The maximum absolute value of all $\Delta Q$
    \item $p_\Delta$: The maximum absolute value of all $\Delta P$
\end{itemize}
These coefficients are incorporated into the model as fixed, non-trainable constants. They are used to rescale both the inputs and outputs: inputs are normalized by the reciprocal of the corresponding coefficient, while outputs are rescaled by multiplication with the coefficients. A schematic illustration of this procedure is shown in Fig.~\ref{fig:data_scalars}. In principle, this ensures that all model inputs lie within the interval $[-1,1]$, and that the outputs remain bounded by the same range. We note, however, that this constraint is not strict, since the model may generate values outside the range spanned by the training data.

\section{Prediction Interval Scaling Analysis}
\label{sec:pred_interval_scaling_analysis}

Longer model prediction intervals decrease the amount of steps - and therefore compute - required to predict a certain trajectory interval. However, a longer prediction interval also naturally decreases the fidelity of the predicted trajectory. While a shorter prediction interval increases the fidelity of the predicted trajectory, there is an increase in compute cost required to predict over a certain length of time.

In figure~\ref{fig:scaling_analysis}, we analyze the prediction error of different models trained with varying prediction intervals. All models predict over the same time interval. As a result, the models with shorter prediction intervals require more prediction steps, and those with longer prediction intervals require fewer prediction steps. All models were trained with data spanning the same overall time interval. Because models with longer prediction intervals naturally require fewer prediction steps to span the same total time interval, the number of trajectory samples in the training dataset of each model varied such that each model was trained with the same amount of overall data. All training data utilized a system size of $L=32$. In the deep quench case, the quench was performed from $g_i = 0.0$ to $g_f = 1.0$. Data was recorded after the trajectory reached its post-transient at $t=0.64$ and continued until $t=2622.08$. In the shallow quench case, the quench was performed from $g_i = 0.5$ to $g_f = 0.8$. Data was recorded from $t=0$ to $t=655.36$. The deep quench case utilized a PARC-based CNN model, whereas the shallow quench case utilized a standard CNN architecture.

In the shallow quench case, the results of figure~\ref{fig:scaling_analysis} demonstrate that all models trained on all prediction intervals are able to effectively perfectly predict a trajectory. Note the extremely small scale of all prediction errors in (a). There is no clear trend in prediction error with respect to the prediction interval of the model used, and any variance could be attributed to simple natural run-to-run variance between models. This is not surprising, given the deterministic nature of the shallow quench case, which we noted as easily captured by a standard CNN in section~\ref{sec:shallow_results}. Note that although we went as high as a prediction interval of $\Delta = 20.48$ in the deep quench case, the maximum prediction interval for the shallow quench case was $\Delta = 10.24$. This was due to the fact that the interval for the deterministic shallow quench case is approximately $\sim 58.8$, and a prediction interval of $20.48$ would be so large as to be ineffective at capturing the overall sinusoidal oscillatory nature of the system.

In the deep quench case, the results of figure~\ref{fig:scaling_analysis} show an increase in prediction error with respect to prediction interval at both extremes of very small prediction intervals and very large prediction intervals. We hypothesize that the increase in prediction error in the small prediction interval regime to be a result of the compounding effects of accumulated error over numerous small steps. In contrast, the increase in prediction error in the large prediction regime may be caused by the simple nature of increasing difficulty of accurate predictions as the prediction interval approaches very large sizes, hindering the ability of the model to accuracy capture or learn the higher-fidelity intricacies of the trajectories. Between the extremes of prediction interval sizes, a middle ground seems to exist which minimized prediction error, likely by balancing the negative effects of both compounding error and large prediction intervals.

\bibliography{ref}

\begin{thebibliography}{62}%
\makeatletter
\providecommand \@ifxundefined [1]{%
 \@ifx{#1\undefined}
}%
\providecommand \@ifnum [1]{%
 \ifnum #1\expandafter \@firstoftwo
 \else \expandafter \@secondoftwo
 \fi
}%
\providecommand \@ifx [1]{%
 \ifx #1\expandafter \@firstoftwo
 \else \expandafter \@secondoftwo
 \fi
}%
\providecommand \natexlab [1]{#1}%
\providecommand \enquote  [1]{``#1''}%
\providecommand \bibnamefont  [1]{#1}%
\providecommand \bibfnamefont [1]{#1}%
\providecommand \citenamefont [1]{#1}%
\providecommand \href@noop [0]{\@secondoftwo}%
\providecommand \href [0]{\begingroup \@sanitize@url \@href}%
\providecommand \@href[1]{\@@startlink{#1}\@@href}%
\providecommand \@@href[1]{\endgroup#1\@@endlink}%
\providecommand \@sanitize@url [0]{\catcode `\\12\catcode `\$12\catcode
  `\&12\catcode `\#12\catcode `\^12\catcode `\_12\catcode `\%12\relax}%
\providecommand \@@startlink[1]{}%
\providecommand \@@endlink[0]{}%
\providecommand \url  [0]{\begingroup\@sanitize@url \@url }%
\providecommand \@url [1]{\endgroup\@href {#1}{\urlprefix }}%
\providecommand \urlprefix  [0]{URL }%
\providecommand \Eprint [0]{\href }%
\providecommand \doibase [0]{https://doi.org/}%
\providecommand \selectlanguage [0]{\@gobble}%
\providecommand \bibinfo  [0]{\@secondoftwo}%
\providecommand \bibfield  [0]{\@secondoftwo}%
\providecommand \translation [1]{[#1]}%
\providecommand \BibitemOpen [0]{}%
\providecommand \bibitemStop [0]{}%
\providecommand \bibitemNoStop [0]{.\EOS\space}%
\providecommand \EOS [0]{\spacefactor3000\relax}%
\providecommand \BibitemShut  [1]{\csname bibitem#1\endcsname}%
\let\auto@bib@innerbib\@empty
%</preamble>
\bibitem [{\citenamefont {Hochreiter}\ and\ \citenamefont
  {Schmidhuber}(1997)}]{hochreiter97}%
  \BibitemOpen
  \bibfield  {author} {\bibinfo {author} {\bibfnamefont {S.}~\bibnamefont
  {Hochreiter}}\ and\ \bibinfo {author} {\bibfnamefont {J.}~\bibnamefont
  {Schmidhuber}},\ }\bibfield  {title} {\bibinfo {title} {Long short-term
  memory},\ }\href {https://doi.org/10.1162/neco.1997.9.8.1735} {\bibfield
  {journal} {\bibinfo  {journal} {Neural Computation}\ }\textbf {\bibinfo
  {volume} {9}},\ \bibinfo {pages} {1735} (\bibinfo {year} {1997})}\BibitemShut
  {NoStop}%
\bibitem [{\citenamefont {Graves}(2012)}]{graves12}%
  \BibitemOpen
  \bibfield  {author} {\bibinfo {author} {\bibfnamefont {A.}~\bibnamefont
  {Graves}},\ }\bibfield  {title} {\bibinfo {title} {Supervised sequence
  labelling},\ }in\ \href@noop {} {\emph {\bibinfo {booktitle} {Supervised Se-
  quence Labelling with Recurrent Neural Networks}}}\ (\bibinfo  {publisher}
  {Springer},\ \bibinfo {address} {New York},\ \bibinfo {year} {2012})\ pp.\
  \bibinfo {pages} {5--13}\BibitemShut {NoStop}%
\bibitem [{\citenamefont {Lipton}\ \emph {et~al.}(2015)\citenamefont {Lipton},
  \citenamefont {Berkowitz},\ and\ \citenamefont {Elkan}}]{lipton15}%
  \BibitemOpen
  \bibfield  {author} {\bibinfo {author} {\bibfnamefont {Z.~C.}\ \bibnamefont
  {Lipton}}, \bibinfo {author} {\bibfnamefont {J.}~\bibnamefont {Berkowitz}},\
  and\ \bibinfo {author} {\bibfnamefont {C.}~\bibnamefont {Elkan}},\ }\href
  {https://arxiv.org/abs/1506.00019} {\bibinfo {title} {A critical review of
  recurrent neural networks for sequence learning}} (\bibinfo {year} {2015}),\
  \Eprint {https://arxiv.org/abs/1506.00019} {arXiv:1506.00019 [cs.LG]}
  \BibitemShut {NoStop}%
\bibitem [{\citenamefont {Sherstinsky}(2020)}]{sherstinsky20}%
  \BibitemOpen
  \bibfield  {author} {\bibinfo {author} {\bibfnamefont {A.}~\bibnamefont
  {Sherstinsky}},\ }\bibfield  {title} {\bibinfo {title} {Fundamentals of
  recurrent neural network (rnn) and long short-term memory (lstm) network},\
  }\href {https://doi.org/https://doi.org/10.1016/j.physd.2019.132306}
  {\bibfield  {journal} {\bibinfo  {journal} {Physica D: Nonlinear Phenomena}\
  }\textbf {\bibinfo {volume} {404}},\ \bibinfo {pages} {132306} (\bibinfo
  {year} {2020})}\BibitemShut {NoStop}%
\bibitem [{\citenamefont {Hu}\ \emph {et~al.}(2020)\citenamefont {Hu},
  \citenamefont {Zhao}, \citenamefont {X{\'u}}, \citenamefont {Lin},\ and\
  \citenamefont {Xu}}]{hu20}%
  \BibitemOpen
  \bibfield  {author} {\bibinfo {author} {\bibfnamefont {Y.}~\bibnamefont
  {Hu}}, \bibinfo {author} {\bibfnamefont {T.}~\bibnamefont {Zhao}}, \bibinfo
  {author} {\bibfnamefont {S.}~\bibnamefont {X{\'u}}}, \bibinfo {author}
  {\bibfnamefont {L.}~\bibnamefont {Lin}},\ and\ \bibinfo {author}
  {\bibfnamefont {Z.}~\bibnamefont {Xu}},\ }\bibfield  {title} {\bibinfo
  {title} {Neural-pde: a rnn based neural network for solving time dependent
  pdes},\ }\href {https://api.semanticscholar.org/CorpusID:239769335}
  {\bibfield  {journal} {\bibinfo  {journal} {Commun. Inf. Syst.}\ }\textbf
  {\bibinfo {volume} {22}},\ \bibinfo {pages} {223} (\bibinfo {year}
  {2020})}\BibitemShut {NoStop}%
\bibitem [{\citenamefont {Karlbauer}\ \emph {et~al.}(2022)\citenamefont
  {Karlbauer}, \citenamefont {Praditia}, \citenamefont {Otte}, \citenamefont
  {Oladyshkin}, \citenamefont {Nowak},\ and\ \citenamefont
  {Butz}}]{karlbauer22}%
  \BibitemOpen
  \bibfield  {author} {\bibinfo {author} {\bibfnamefont {M.}~\bibnamefont
  {Karlbauer}}, \bibinfo {author} {\bibfnamefont {T.}~\bibnamefont {Praditia}},
  \bibinfo {author} {\bibfnamefont {S.}~\bibnamefont {Otte}}, \bibinfo {author}
  {\bibfnamefont {S.}~\bibnamefont {Oladyshkin}}, \bibinfo {author}
  {\bibfnamefont {W.}~\bibnamefont {Nowak}},\ and\ \bibinfo {author}
  {\bibfnamefont {M.~V.}\ \bibnamefont {Butz}},\ }\bibfield  {title} {\bibinfo
  {title} {Composing partial differential equations with physics-aware neural
  networks},\ }in\ \href@noop {} {\emph {\bibinfo {booktitle} {International
  Conference on Machine Learning}}}\ (\bibinfo {organization} {PMLR},\ \bibinfo
  {year} {2022})\ pp.\ \bibinfo {pages} {10773--10801}\BibitemShut {NoStop}%
\bibitem [{\citenamefont {Hafiz}\ \emph {et~al.}(2024)\citenamefont {Hafiz},
  \citenamefont {Faiq},\ and\ \citenamefont {Hassaballah}}]{hafiz24}%
  \BibitemOpen
  \bibfield  {author} {\bibinfo {author} {\bibfnamefont {A.~M.}\ \bibnamefont
  {Hafiz}}, \bibinfo {author} {\bibfnamefont {I.}~\bibnamefont {Faiq}},\ and\
  \bibinfo {author} {\bibfnamefont {M.}~\bibnamefont {Hassaballah}},\
  }\bibfield  {title} {\bibinfo {title} {Solving partial differential equations
  using large-data models: a literature review},\ }\href
  {https://doi.org/10.1007/s10462-024-10784-5} {\bibfield  {journal} {\bibinfo
  {journal} {Artificial Intelligence Review}\ }\textbf {\bibinfo {volume}
  {57}},\ \bibinfo {pages} {152} (\bibinfo {year} {2024})}\BibitemShut
  {NoStop}%
\bibitem [{\citenamefont {Wu}\ \emph {et~al.}(2022)\citenamefont {Wu},
  \citenamefont {Hennigh}, \citenamefont {Kautz}, \citenamefont {Choudhry},\
  and\ \citenamefont {Byeon}}]{wu22}%
  \BibitemOpen
  \bibfield  {author} {\bibinfo {author} {\bibfnamefont {B.}~\bibnamefont
  {Wu}}, \bibinfo {author} {\bibfnamefont {O.}~\bibnamefont {Hennigh}},
  \bibinfo {author} {\bibfnamefont {J.}~\bibnamefont {Kautz}}, \bibinfo
  {author} {\bibfnamefont {S.}~\bibnamefont {Choudhry}},\ and\ \bibinfo
  {author} {\bibfnamefont {W.}~\bibnamefont {Byeon}},\ }\bibfield  {title}
  {\bibinfo {title} {Physics informed rnn-dct networks for time-dependent
  partial differential equations},\ }in\ \href@noop {} {\emph {\bibinfo
  {booktitle} {Computational Science -- ICCS 2022}}},\ \bibinfo {editor}
  {edited by\ \bibinfo {editor} {\bibfnamefont {D.}~\bibnamefont {Groen}},
  \bibinfo {editor} {\bibfnamefont {C.}~\bibnamefont {de~Mulatier}}, \bibinfo
  {editor} {\bibfnamefont {M.}~\bibnamefont {Paszynski}}, \bibinfo {editor}
  {\bibfnamefont {V.~V.}\ \bibnamefont {Krzhizhanovskaya}}, \bibinfo {editor}
  {\bibfnamefont {J.~J.}\ \bibnamefont {Dongarra}},\ and\ \bibinfo {editor}
  {\bibfnamefont {P.~M.~A.}\ \bibnamefont {Sloot}}}\ (\bibinfo  {publisher}
  {Springer International Publishing},\ \bibinfo {address} {Cham},\ \bibinfo
  {year} {2022})\ pp.\ \bibinfo {pages} {372--379}\BibitemShut {NoStop}%
\bibitem [{\citenamefont {Sun}\ \emph {et~al.}(2020)\citenamefont {Sun},
  \citenamefont {Zhang},\ and\ \citenamefont {Schaeffer}}]{sun20}%
  \BibitemOpen
  \bibfield  {author} {\bibinfo {author} {\bibfnamefont {Y.}~\bibnamefont
  {Sun}}, \bibinfo {author} {\bibfnamefont {L.}~\bibnamefont {Zhang}},\ and\
  \bibinfo {author} {\bibfnamefont {H.}~\bibnamefont {Schaeffer}},\ }\bibfield
  {title} {\bibinfo {title} {{NeuPDE: N}eural network based ordinary and
  partial differential equations for modeling time-dependent data},\ }in\ \href
  {https://proceedings.mlr.press/v107/sun20a.html} {\emph {\bibinfo {booktitle}
  {Proceedings of The First Mathematical and Scientific Machine Learning
  Conference}}},\ \bibinfo {series} {Proceedings of Machine Learning Research},
  Vol.\ \bibinfo {volume} {107},\ \bibinfo {editor} {edited by\ \bibinfo
  {editor} {\bibfnamefont {J.}~\bibnamefont {Lu}}\ and\ \bibinfo {editor}
  {\bibfnamefont {R.}~\bibnamefont {Ward}}}\ (\bibinfo  {publisher} {PMLR},\
  \bibinfo {year} {2020})\ pp.\ \bibinfo {pages} {352--372}\BibitemShut
  {NoStop}%
\bibitem [{\citenamefont {Zhong}\ \emph {et~al.}(2021)\citenamefont {Zhong},
  \citenamefont {Sun},\ and\ \citenamefont {Qiu}}]{zhong21}%
  \BibitemOpen
  \bibfield  {author} {\bibinfo {author} {\bibfnamefont {Q.}~\bibnamefont
  {Zhong}}, \bibinfo {author} {\bibfnamefont {Y.}~\bibnamefont {Sun}},\ and\
  \bibinfo {author} {\bibfnamefont {Z.}~\bibnamefont {Qiu}},\ }\bibfield
  {title} {\bibinfo {title} {Deep recurrent neural network with sharing weights
  for solving high-dimensional pdes},\ }in\ \href
  {https://doi.org/10.1109/ICDSCA53499.2021.9650325} {\emph {\bibinfo
  {booktitle} {2021 IEEE International Conference on Data Science and Computer
  Application (ICDSCA)}}}\ (\bibinfo {year} {2021})\ pp.\ \bibinfo {pages}
  {6--9}\BibitemShut {NoStop}%
\bibitem [{\citenamefont {Liang}\ \emph {et~al.}(2024)\citenamefont {Liang},
  \citenamefont {Niu}, \citenamefont {Yue},\ and\ \citenamefont
  {Lei}}]{liang24}%
  \BibitemOpen
  \bibfield  {author} {\bibinfo {author} {\bibfnamefont {Y.}~\bibnamefont
  {Liang}}, \bibinfo {author} {\bibfnamefont {R.}~\bibnamefont {Niu}}, \bibinfo
  {author} {\bibfnamefont {J.}~\bibnamefont {Yue}},\ and\ \bibinfo {author}
  {\bibfnamefont {M.}~\bibnamefont {Lei}},\ }\bibfield  {title} {\bibinfo
  {title} {A physics-informed recurrent neural network for solving
  time-dependent partial differential equations},\ }\href
  {https://doi.org/10.1142/S0219876223410037} {\bibfield  {journal} {\bibinfo
  {journal} {International Journal of Computational Methods}\ }\textbf
  {\bibinfo {volume} {21}},\ \bibinfo {pages} {2341003} (\bibinfo {year}
  {2024})}\BibitemShut {NoStop}%
\bibitem [{\citenamefont {Long}\ \emph {et~al.}(2018)\citenamefont {Long},
  \citenamefont {Lu}, \citenamefont {Ma},\ and\ \citenamefont {Dong}}]{long18}%
  \BibitemOpen
  \bibfield  {author} {\bibinfo {author} {\bibfnamefont {Z.}~\bibnamefont
  {Long}}, \bibinfo {author} {\bibfnamefont {Y.}~\bibnamefont {Lu}}, \bibinfo
  {author} {\bibfnamefont {X.}~\bibnamefont {Ma}},\ and\ \bibinfo {author}
  {\bibfnamefont {B.}~\bibnamefont {Dong}},\ }\bibfield  {title} {\bibinfo
  {title} {{PDE}-net: Learning {PDE}s from data},\ }in\ \href
  {https://proceedings.mlr.press/v80/long18a.html} {\emph {\bibinfo {booktitle}
  {Proceedings of the 35th International Conference on Machine Learning}}},\
  \bibinfo {series} {Proceedings of Machine Learning Research}, Vol.~\bibinfo
  {volume} {80},\ \bibinfo {editor} {edited by\ \bibinfo {editor}
  {\bibfnamefont {J.}~\bibnamefont {Dy}}\ and\ \bibinfo {editor} {\bibfnamefont
  {A.}~\bibnamefont {Krause}}}\ (\bibinfo  {publisher} {PMLR},\ \bibinfo {year}
  {2018})\ pp.\ \bibinfo {pages} {3208--3216}\BibitemShut {NoStop}%
\bibitem [{\citenamefont {Long}\ \emph {et~al.}(2019)\citenamefont {Long},
  \citenamefont {Lu},\ and\ \citenamefont {Dong}}]{long19}%
  \BibitemOpen
  \bibfield  {author} {\bibinfo {author} {\bibfnamefont {Z.}~\bibnamefont
  {Long}}, \bibinfo {author} {\bibfnamefont {Y.}~\bibnamefont {Lu}},\ and\
  \bibinfo {author} {\bibfnamefont {B.}~\bibnamefont {Dong}},\ }\bibfield
  {title} {\bibinfo {title} {Pde-net 2.0: Learning pdes from data with a
  numeric-symbolic hybrid deep network},\ }\href
  {https://doi.org/https://doi.org/10.1016/j.jcp.2019.108925} {\bibfield
  {journal} {\bibinfo  {journal} {Journal of Computational Physics}\ }\textbf
  {\bibinfo {volume} {399}},\ \bibinfo {pages} {108925} (\bibinfo {year}
  {2019})}\BibitemShut {NoStop}%
\bibitem [{\citenamefont {Zuo}\ \emph {et~al.}(2015)\citenamefont {Zuo},
  \citenamefont {Shuai}, \citenamefont {Wang}, \citenamefont {Liu},
  \citenamefont {Wang}, \citenamefont {Wang},\ and\ \citenamefont
  {Chen}}]{zuo15}%
  \BibitemOpen
  \bibfield  {author} {\bibinfo {author} {\bibfnamefont {Z.}~\bibnamefont
  {Zuo}}, \bibinfo {author} {\bibfnamefont {B.}~\bibnamefont {Shuai}}, \bibinfo
  {author} {\bibfnamefont {G.}~\bibnamefont {Wang}}, \bibinfo {author}
  {\bibfnamefont {X.}~\bibnamefont {Liu}}, \bibinfo {author} {\bibfnamefont
  {X.}~\bibnamefont {Wang}}, \bibinfo {author} {\bibfnamefont {B.}~\bibnamefont
  {Wang}},\ and\ \bibinfo {author} {\bibfnamefont {Y.}~\bibnamefont {Chen}},\
  }\bibfield  {title} {\bibinfo {title} {Convolutional recurrent neural
  networks: Learning spatial dependencies for image representation},\ }in\
  \href {https://doi.org/10.1109/CVPRW.2015.7301268} {\emph {\bibinfo
  {booktitle} {2015 IEEE Conference on Computer Vision and Pattern Recognition
  Workshops (CVPRW)}}}\ (\bibinfo {year} {2015})\ pp.\ \bibinfo {pages}
  {18--26}\BibitemShut {NoStop}%
\bibitem [{\citenamefont {Saha}\ \emph {et~al.}(2021)\citenamefont {Saha},
  \citenamefont {Dash},\ and\ \citenamefont {Mukhopadhyay}}]{saha21}%
  \BibitemOpen
  \bibfield  {author} {\bibinfo {author} {\bibfnamefont {P.}~\bibnamefont
  {Saha}}, \bibinfo {author} {\bibfnamefont {S.}~\bibnamefont {Dash}},\ and\
  \bibinfo {author} {\bibfnamefont {S.}~\bibnamefont {Mukhopadhyay}},\
  }\bibfield  {title} {\bibinfo {title} {Physics-incorporated convolutional
  recurrent neural networks for source identification and forecasting of
  dynamical systems},\ }\href
  {https://doi.org/https://doi.org/10.1016/j.neunet.2021.08.033} {\bibfield
  {journal} {\bibinfo  {journal} {Neural Networks}\ }\textbf {\bibinfo {volume}
  {144}},\ \bibinfo {pages} {359} (\bibinfo {year} {2021})}\BibitemShut
  {NoStop}%
\bibitem [{\citenamefont {Ren}\ \emph {et~al.}(2022)\citenamefont {Ren},
  \citenamefont {Rao}, \citenamefont {Liu}, \citenamefont {Wang},\ and\
  \citenamefont {Sun}}]{ren22}%
  \BibitemOpen
  \bibfield  {author} {\bibinfo {author} {\bibfnamefont {P.}~\bibnamefont
  {Ren}}, \bibinfo {author} {\bibfnamefont {C.}~\bibnamefont {Rao}}, \bibinfo
  {author} {\bibfnamefont {Y.}~\bibnamefont {Liu}}, \bibinfo {author}
  {\bibfnamefont {J.-X.}\ \bibnamefont {Wang}},\ and\ \bibinfo {author}
  {\bibfnamefont {H.}~\bibnamefont {Sun}},\ }\bibfield  {title} {\bibinfo
  {title} {Phycrnet: Physics-informed convolutional-recurrent network for
  solving spatiotemporal pdes},\ }\href
  {https://doi.org/https://doi.org/10.1016/j.cma.2021.114399} {\bibfield
  {journal} {\bibinfo  {journal} {Computer Methods in Applied Mechanics and
  Engineering}\ }\textbf {\bibinfo {volume} {389}},\ \bibinfo {pages} {114399}
  (\bibinfo {year} {2022})}\BibitemShut {NoStop}%
\bibitem [{\citenamefont {Nguyen}\ \emph {et~al.}(2023)\citenamefont {Nguyen},
  \citenamefont {Nguyen}, \citenamefont {Choi}, \citenamefont {Seshadri},
  \citenamefont {Udaykumar},\ and\ \citenamefont {Baek}}]{nguyen23}%
  \BibitemOpen
  \bibfield  {author} {\bibinfo {author} {\bibfnamefont {P.~C.}\ \bibnamefont
  {Nguyen}}, \bibinfo {author} {\bibfnamefont {Y.-T.}\ \bibnamefont {Nguyen}},
  \bibinfo {author} {\bibfnamefont {J.~B.}\ \bibnamefont {Choi}}, \bibinfo
  {author} {\bibfnamefont {P.~K.}\ \bibnamefont {Seshadri}}, \bibinfo {author}
  {\bibfnamefont {H.~S.}\ \bibnamefont {Udaykumar}},\ and\ \bibinfo {author}
  {\bibfnamefont {S.~S.}\ \bibnamefont {Baek}},\ }\bibfield  {title} {\bibinfo
  {title} {{PARC}: Physics-aware recurrent convolutional neural networks to
  assimilate meso scale reactive mechanics of energetic materials},\ }\href
  {https://doi.org/10.1126/sciadv.add6868} {\bibfield  {journal} {\bibinfo
  {journal} {Science Advances}\ }\textbf {\bibinfo {volume} {9}},\ \bibinfo
  {pages} {eadd6868} (\bibinfo {year} {2023})}\BibitemShut {NoStop}%
\bibitem [{\citenamefont {Nguyen}\ \emph {et~al.}(2024)\citenamefont {Nguyen},
  \citenamefont {Cheng}, \citenamefont {Azarfar}, \citenamefont {Seshadri},
  \citenamefont {Nguyen}, \citenamefont {Kim}, \citenamefont {Choi},
  \citenamefont {Udaykumar},\ and\ \citenamefont {Baek}}]{nguyen24}%
  \BibitemOpen
  \bibfield  {author} {\bibinfo {author} {\bibfnamefont {P.~C.~H.}\
  \bibnamefont {Nguyen}}, \bibinfo {author} {\bibfnamefont {X.}~\bibnamefont
  {Cheng}}, \bibinfo {author} {\bibfnamefont {S.}~\bibnamefont {Azarfar}},
  \bibinfo {author} {\bibfnamefont {P.}~\bibnamefont {Seshadri}}, \bibinfo
  {author} {\bibfnamefont {Y.~T.}\ \bibnamefont {Nguyen}}, \bibinfo {author}
  {\bibfnamefont {M.}~\bibnamefont {Kim}}, \bibinfo {author} {\bibfnamefont
  {S.}~\bibnamefont {Choi}}, \bibinfo {author} {\bibfnamefont {H.~S.}\
  \bibnamefont {Udaykumar}},\ and\ \bibinfo {author} {\bibfnamefont
  {S.}~\bibnamefont {Baek}},\ }\href {https://arxiv.org/abs/2402.12503}
  {\bibinfo {title} {{PARCv2}: Physics-aware recurrent convolutional neural
  networks for spatiotemporal dynamics modeling}} (\bibinfo {year} {2024}),\
  \Eprint {https://arxiv.org/abs/2402.12503} {arXiv:2402.12503 [cs.LG]}
  \BibitemShut {NoStop}%
\bibitem [{\citenamefont {Press}\ \emph {et~al.}(1992)\citenamefont {Press},
  \citenamefont {Teukolsky}, \citenamefont {Vetterling},\ and\ \citenamefont
  {Flannery}}]{press92}%
  \BibitemOpen
  \bibfield  {author} {\bibinfo {author} {\bibfnamefont {W.~H.}\ \bibnamefont
  {Press}}, \bibinfo {author} {\bibfnamefont {S.~A.}\ \bibnamefont
  {Teukolsky}}, \bibinfo {author} {\bibfnamefont {W.~T.}\ \bibnamefont
  {Vetterling}},\ and\ \bibinfo {author} {\bibfnamefont {B.~P.}\ \bibnamefont
  {Flannery}},\ }\href@noop {} {\emph {\bibinfo {title} {Numerical Recipes in
  C}}},\ \bibinfo {edition} {2nd}\ ed.\ (\bibinfo  {publisher} {Cambridge
  University Press},\ \bibinfo {address} {Cambridge, USA},\ \bibinfo {year}
  {1992})\BibitemShut {NoStop}%
\bibitem [{\citenamefont {Morton}\ and\ \citenamefont
  {Mayers}(2005)}]{morton05}%
  \BibitemOpen
  \bibfield  {author} {\bibinfo {author} {\bibfnamefont {K.~W.}\ \bibnamefont
  {Morton}}\ and\ \bibinfo {author} {\bibfnamefont {D.~F.}\ \bibnamefont
  {Mayers}},\ }\href@noop {} {\emph {\bibinfo {title} {Numerical Solution of
  Partial Differential Equations: An Introduction}}},\ \bibinfo {edition}
  {2nd}\ ed.\ (\bibinfo  {publisher} {Cambridge University Press},\ \bibinfo
  {year} {2005})\BibitemShut {NoStop}%
\bibitem [{\citenamefont {Ascher}\ \emph {et~al.}(1997)\citenamefont {Ascher},
  \citenamefont {Ruuth},\ and\ \citenamefont {Spiteri}}]{ascher97}%
  \BibitemOpen
  \bibfield  {author} {\bibinfo {author} {\bibfnamefont {U.~M.}\ \bibnamefont
  {Ascher}}, \bibinfo {author} {\bibfnamefont {S.~J.}\ \bibnamefont {Ruuth}},\
  and\ \bibinfo {author} {\bibfnamefont {R.~J.}\ \bibnamefont {Spiteri}},\
  }\bibfield  {title} {\bibinfo {title} {Implicit-explicit runge-kutta methods
  for time-dependent partial differential equations},\ }\href
  {https://doi.org/https://doi.org/10.1016/S0168-9274(97)00056-1} {\bibfield
  {journal} {\bibinfo  {journal} {Applied Numerical Mathematics}\ }\textbf
  {\bibinfo {volume} {25}},\ \bibinfo {pages} {151} (\bibinfo {year} {1997})},\
  \bibinfo {note} {special Issue on Time Integration}\BibitemShut {NoStop}%
\bibitem [{\citenamefont {Pathak}\ \emph {et~al.}(2017)\citenamefont {Pathak},
  \citenamefont {Lu}, \citenamefont {Hunt}, \citenamefont {Girvan},\ and\
  \citenamefont {Ott}}]{Pathak_2017}%
  \BibitemOpen
  \bibfield  {author} {\bibinfo {author} {\bibfnamefont {J.}~\bibnamefont
  {Pathak}}, \bibinfo {author} {\bibfnamefont {Z.}~\bibnamefont {Lu}}, \bibinfo
  {author} {\bibfnamefont {B.~R.}\ \bibnamefont {Hunt}}, \bibinfo {author}
  {\bibfnamefont {M.}~\bibnamefont {Girvan}},\ and\ \bibinfo {author}
  {\bibfnamefont {E.}~\bibnamefont {Ott}},\ }\bibfield  {title} {\bibinfo
  {title} {Using machine learning to replicate chaotic attractors and calculate
  lyapunov exponents from data},\ }\bibfield  {journal} {\bibinfo  {journal}
  {Chaos: An Interdisciplinary Journal of Nonlinear Science}\ }\textbf
  {\bibinfo {volume} {27}},\ \href {https://doi.org/10.1063/1.5010300}
  {10.1063/1.5010300} (\bibinfo {year} {2017})\BibitemShut {NoStop}%
\bibitem [{\citenamefont {Haluszczynski}\ and\ \citenamefont
  {Räth}(2019)}]{Haluszczynski_2019}%
  \BibitemOpen
  \bibfield  {author} {\bibinfo {author} {\bibfnamefont {A.}~\bibnamefont
  {Haluszczynski}}\ and\ \bibinfo {author} {\bibfnamefont {C.}~\bibnamefont
  {Räth}},\ }\bibfield  {title} {\bibinfo {title} {Good and bad predictions:
  Assessing and improving the replication of chaotic attractors by means of
  reservoir computing},\ }\bibfield  {journal} {\bibinfo  {journal} {Chaos: An
  Interdisciplinary Journal of Nonlinear Science}\ }\textbf {\bibinfo {volume}
  {29}},\ \href {https://doi.org/10.1063/1.5118725} {10.1063/1.5118725}
  (\bibinfo {year} {2019})\BibitemShut {NoStop}%
\bibitem [{\citenamefont {Pathak}\ \emph {et~al.}(2018)\citenamefont {Pathak},
  \citenamefont {Hunt}, \citenamefont {Girvan}, \citenamefont {Lu},\ and\
  \citenamefont {Ott}}]{PhysRevLett.120.024102}%
  \BibitemOpen
  \bibfield  {author} {\bibinfo {author} {\bibfnamefont {J.}~\bibnamefont
  {Pathak}}, \bibinfo {author} {\bibfnamefont {B.}~\bibnamefont {Hunt}},
  \bibinfo {author} {\bibfnamefont {M.}~\bibnamefont {Girvan}}, \bibinfo
  {author} {\bibfnamefont {Z.}~\bibnamefont {Lu}},\ and\ \bibinfo {author}
  {\bibfnamefont {E.}~\bibnamefont {Ott}},\ }\bibfield  {title} {\bibinfo
  {title} {Model-free prediction of large spatiotemporally chaotic systems from
  data: A reservoir computing approach},\ }\href
  {https://doi.org/10.1103/PhysRevLett.120.024102} {\bibfield  {journal}
  {\bibinfo  {journal} {Phys. Rev. Lett.}\ }\textbf {\bibinfo {volume} {120}},\
  \bibinfo {pages} {024102} (\bibinfo {year} {2018})}\BibitemShut {NoStop}%
\bibitem [{\citenamefont {Pershin}\ \emph {et~al.}(2023)\citenamefont
  {Pershin}, \citenamefont {Beaume}, \citenamefont {Li},\ and\ \citenamefont
  {Tobias}}]{PhysRevE.107.014304}%
  \BibitemOpen
  \bibfield  {author} {\bibinfo {author} {\bibfnamefont {A.}~\bibnamefont
  {Pershin}}, \bibinfo {author} {\bibfnamefont {C.}~\bibnamefont {Beaume}},
  \bibinfo {author} {\bibfnamefont {K.}~\bibnamefont {Li}},\ and\ \bibinfo
  {author} {\bibfnamefont {S.~M.}\ \bibnamefont {Tobias}},\ }\bibfield  {title}
  {\bibinfo {title} {Training a neural network to predict dynamics it has never
  seen},\ }\href {https://doi.org/10.1103/PhysRevE.107.014304} {\bibfield
  {journal} {\bibinfo  {journal} {Phys. Rev. E}\ }\textbf {\bibinfo {volume}
  {107}},\ \bibinfo {pages} {014304} (\bibinfo {year} {2023})}\BibitemShut
  {NoStop}%
\bibitem [{\citenamefont {Mandal}\ \emph {et~al.}(2022)\citenamefont {Mandal},
  \citenamefont {Sinha},\ and\ \citenamefont {Shrimali}}]{PhysRevE.105.054203}%
  \BibitemOpen
  \bibfield  {author} {\bibinfo {author} {\bibfnamefont {S.}~\bibnamefont
  {Mandal}}, \bibinfo {author} {\bibfnamefont {S.}~\bibnamefont {Sinha}},\ and\
  \bibinfo {author} {\bibfnamefont {M.~D.}\ \bibnamefont {Shrimali}},\
  }\bibfield  {title} {\bibinfo {title} {Machine-learning potential of a single
  pendulum},\ }\href {https://doi.org/10.1103/PhysRevE.105.054203} {\bibfield
  {journal} {\bibinfo  {journal} {Phys. Rev. E}\ }\textbf {\bibinfo {volume}
  {105}},\ \bibinfo {pages} {054203} (\bibinfo {year} {2022})}\BibitemShut
  {NoStop}%
\bibitem [{\citenamefont {Goodfellow}\ \emph
  {et~al.}(2016{\natexlab{a}})\citenamefont {Goodfellow}, \citenamefont
  {Bengio},\ and\ \citenamefont {Courville}}]{goodfellow16}%
  \BibitemOpen
  \bibfield  {author} {\bibinfo {author} {\bibfnamefont {I.}~\bibnamefont
  {Goodfellow}}, \bibinfo {author} {\bibfnamefont {Y.}~\bibnamefont {Bengio}},\
  and\ \bibinfo {author} {\bibfnamefont {A.}~\bibnamefont {Courville}},\
  }\bibfield  {title} {\bibinfo {title} {Convolutional networks},\ }in\ \href
  {http://www.deeplearningbook.org} {\emph {\bibinfo {booktitle} {Deep
  Learning}}}\ (\bibinfo  {publisher} {MIT Press},\ \bibinfo {year} {2016})\
  Chap.~\bibinfo {chapter} {9}, pp.\ \bibinfo {pages} {326--366}\BibitemShut
  {NoStop}%
\bibitem [{\citenamefont {Behler}\ and\ \citenamefont
  {Parrinello}(2007)}]{behler07}%
  \BibitemOpen
  \bibfield  {author} {\bibinfo {author} {\bibfnamefont {J.}~\bibnamefont
  {Behler}}\ and\ \bibinfo {author} {\bibfnamefont {M.}~\bibnamefont
  {Parrinello}},\ }\bibfield  {title} {\bibinfo {title} {Generalized
  neural-network representation of high-dimensional potential-energy
  surfaces},\ }\href {https://doi.org/10.1103/PhysRevLett.98.146401} {\bibfield
   {journal} {\bibinfo  {journal} {Phys. Rev. Lett.}\ }\textbf {\bibinfo
  {volume} {98}},\ \bibinfo {pages} {146401} (\bibinfo {year}
  {2007})}\BibitemShut {NoStop}%
\bibitem [{\citenamefont {Bart\'ok}\ \emph {et~al.}(2010)\citenamefont
  {Bart\'ok}, \citenamefont {Payne}, \citenamefont {Kondor},\ and\
  \citenamefont {Cs\'anyi}}]{bartok10}%
  \BibitemOpen
  \bibfield  {author} {\bibinfo {author} {\bibfnamefont {A.~P.}\ \bibnamefont
  {Bart\'ok}}, \bibinfo {author} {\bibfnamefont {M.~C.}\ \bibnamefont {Payne}},
  \bibinfo {author} {\bibfnamefont {R.}~\bibnamefont {Kondor}},\ and\ \bibinfo
  {author} {\bibfnamefont {G.}~\bibnamefont {Cs\'anyi}},\ }\bibfield  {title}
  {\bibinfo {title} {Gaussian approximation potentials: The accuracy of quantum
  mechanics, without the electrons},\ }\href
  {https://doi.org/10.1103/PhysRevLett.104.136403} {\bibfield  {journal}
  {\bibinfo  {journal} {Phys. Rev. Lett.}\ }\textbf {\bibinfo {volume} {104}},\
  \bibinfo {pages} {136403} (\bibinfo {year} {2010})}\BibitemShut {NoStop}%
\bibitem [{\citenamefont {Li}\ \emph {et~al.}(2015)\citenamefont {Li},
  \citenamefont {Kermode},\ and\ \citenamefont {De~Vita}}]{li15}%
  \BibitemOpen
  \bibfield  {author} {\bibinfo {author} {\bibfnamefont {Z.}~\bibnamefont
  {Li}}, \bibinfo {author} {\bibfnamefont {J.~R.}\ \bibnamefont {Kermode}},\
  and\ \bibinfo {author} {\bibfnamefont {A.}~\bibnamefont {De~Vita}},\
  }\bibfield  {title} {\bibinfo {title} {Molecular dynamics with on-the-fly
  machine learning of quantum-mechanical forces},\ }\href
  {https://doi.org/10.1103/PhysRevLett.114.096405} {\bibfield  {journal}
  {\bibinfo  {journal} {Phys. Rev. Lett.}\ }\textbf {\bibinfo {volume} {114}},\
  \bibinfo {pages} {096405} (\bibinfo {year} {2015})}\BibitemShut {NoStop}%
\bibitem [{\citenamefont {Shapeev}(2016)}]{shapeev16}%
  \BibitemOpen
  \bibfield  {author} {\bibinfo {author} {\bibfnamefont {A.~V.}\ \bibnamefont
  {Shapeev}},\ }\bibfield  {title} {\bibinfo {title} {Moment tensor potentials:
  A class of systematically improvable interatomic potentials},\ }\href
  {https://doi.org/10.1137/15M1054183} {\bibfield  {journal} {\bibinfo
  {journal} {Multiscale Modeling \& Simulation}\ }\textbf {\bibinfo {volume}
  {14}},\ \bibinfo {pages} {1153} (\bibinfo {year} {2016})}\BibitemShut
  {NoStop}%
\bibitem [{\citenamefont {Botu}\ \emph {et~al.}(2017)\citenamefont {Botu},
  \citenamefont {Batra}, \citenamefont {Chapman},\ and\ \citenamefont
  {Ramprasad}}]{botu17}%
  \BibitemOpen
  \bibfield  {author} {\bibinfo {author} {\bibfnamefont {V.}~\bibnamefont
  {Botu}}, \bibinfo {author} {\bibfnamefont {R.}~\bibnamefont {Batra}},
  \bibinfo {author} {\bibfnamefont {J.}~\bibnamefont {Chapman}},\ and\ \bibinfo
  {author} {\bibfnamefont {R.}~\bibnamefont {Ramprasad}},\ }\bibfield  {title}
  {\bibinfo {title} {Machine learning force fields: Construction, validation,
  and outlook},\ }\href {https://doi.org/10.1021/acs.jpcc.6b10908} {\bibfield
  {journal} {\bibinfo  {journal} {The Journal of Physical Chemistry C}\
  }\textbf {\bibinfo {volume} {121}},\ \bibinfo {pages} {511} (\bibinfo {year}
  {2017})}\BibitemShut {NoStop}%
\bibitem [{\citenamefont {Chmiela}\ \emph {et~al.}(2017)\citenamefont
  {Chmiela}, \citenamefont {Tkatchenko}, \citenamefont {Sauceda}, \citenamefont
  {Poltavsky}, \citenamefont {Schütt},\ and\ \citenamefont
  {Müller}}]{chmiela17}%
  \BibitemOpen
  \bibfield  {author} {\bibinfo {author} {\bibfnamefont {S.}~\bibnamefont
  {Chmiela}}, \bibinfo {author} {\bibfnamefont {A.}~\bibnamefont {Tkatchenko}},
  \bibinfo {author} {\bibfnamefont {H.~E.}\ \bibnamefont {Sauceda}}, \bibinfo
  {author} {\bibfnamefont {I.}~\bibnamefont {Poltavsky}}, \bibinfo {author}
  {\bibfnamefont {K.~T.}\ \bibnamefont {Schütt}},\ and\ \bibinfo {author}
  {\bibfnamefont {K.-R.}\ \bibnamefont {Müller}},\ }\bibfield  {title}
  {\bibinfo {title} {Machine learning of accurate energy-conserving molecular
  force fields},\ }\href {https://doi.org/10.1126/sciadv.1603015} {\bibfield
  {journal} {\bibinfo  {journal} {Science Advances}\ }\textbf {\bibinfo
  {volume} {3}},\ \bibinfo {pages} {e1603015} (\bibinfo {year}
  {2017})}\BibitemShut {NoStop}%
\bibitem [{\citenamefont {Chmiela}\ \emph {et~al.}(2018)\citenamefont
  {Chmiela}, \citenamefont {Sauceda}, \citenamefont {M{\"u}ller},\ and\
  \citenamefont {Tkatchenko}}]{chmiela18}%
  \BibitemOpen
  \bibfield  {author} {\bibinfo {author} {\bibfnamefont {S.}~\bibnamefont
  {Chmiela}}, \bibinfo {author} {\bibfnamefont {H.~E.}\ \bibnamefont
  {Sauceda}}, \bibinfo {author} {\bibfnamefont {K.-R.}\ \bibnamefont
  {M{\"u}ller}},\ and\ \bibinfo {author} {\bibfnamefont {A.}~\bibnamefont
  {Tkatchenko}},\ }\bibfield  {title} {\bibinfo {title} {Towards exact
  molecular dynamics simulations with machine-learned force fields},\ }\href
  {https://doi.org/10.1038/s41467-018-06169-2} {\bibfield  {journal} {\bibinfo
  {journal} {Nature Communications}\ }\textbf {\bibinfo {volume} {9}},\
  \bibinfo {pages} {3887} (\bibinfo {year} {2018})}\BibitemShut {NoStop}%
\bibitem [{\citenamefont {Weiler}\ \emph {et~al.}(2018)\citenamefont {Weiler},
  \citenamefont {Geiger}, \citenamefont {Welling}, \citenamefont {Boomsma},\
  and\ \citenamefont {Cohen}}]{weiler18}%
  \BibitemOpen
  \bibfield  {author} {\bibinfo {author} {\bibfnamefont {M.}~\bibnamefont
  {Weiler}}, \bibinfo {author} {\bibfnamefont {M.}~\bibnamefont {Geiger}},
  \bibinfo {author} {\bibfnamefont {M.}~\bibnamefont {Welling}}, \bibinfo
  {author} {\bibfnamefont {W.}~\bibnamefont {Boomsma}},\ and\ \bibinfo {author}
  {\bibfnamefont {T.~S.}\ \bibnamefont {Cohen}},\ }\bibfield  {title} {\bibinfo
  {title} {3d steerable cnns: Learning rotationally equivariant features in
  volumetric data},\ }\href@noop {} {\bibfield  {journal} {\bibinfo  {journal}
  {Advances in Neural Information Processing Systems}\ }\textbf {\bibinfo
  {volume} {31}} (\bibinfo {year} {2018})}\BibitemShut {NoStop}%
\bibitem [{\citenamefont {Batzner}\ \emph {et~al.}(2022)\citenamefont
  {Batzner}, \citenamefont {Musaelian}, \citenamefont {Sun}, \citenamefont
  {Geiger}, \citenamefont {Mailoa}, \citenamefont {Kornbluth}, \citenamefont
  {Molinari}, \citenamefont {Smidt},\ and\ \citenamefont
  {Kozinsky}}]{batzner22}%
  \BibitemOpen
  \bibfield  {author} {\bibinfo {author} {\bibfnamefont {S.}~\bibnamefont
  {Batzner}}, \bibinfo {author} {\bibfnamefont {A.}~\bibnamefont {Musaelian}},
  \bibinfo {author} {\bibfnamefont {L.}~\bibnamefont {Sun}}, \bibinfo {author}
  {\bibfnamefont {M.}~\bibnamefont {Geiger}}, \bibinfo {author} {\bibfnamefont
  {J.~P.}\ \bibnamefont {Mailoa}}, \bibinfo {author} {\bibfnamefont
  {M.}~\bibnamefont {Kornbluth}}, \bibinfo {author} {\bibfnamefont
  {N.}~\bibnamefont {Molinari}}, \bibinfo {author} {\bibfnamefont {T.~E.}\
  \bibnamefont {Smidt}},\ and\ \bibinfo {author} {\bibfnamefont
  {B.}~\bibnamefont {Kozinsky}},\ }\bibfield  {title} {\bibinfo {title}
  {E(3)-equivariant graph neural networks for data-efficient and accurate
  interatomic potentials},\ }\href {https://doi.org/10.1038/s41467-022-29939-5}
  {\bibfield  {journal} {\bibinfo  {journal} {Nature Communications}\ }\textbf
  {\bibinfo {volume} {13}},\ \bibinfo {pages} {2453} (\bibinfo {year}
  {2022})}\BibitemShut {NoStop}%
\bibitem [{\citenamefont {Gong}\ \emph {et~al.}(2023)\citenamefont {Gong},
  \citenamefont {Li}, \citenamefont {Zou}, \citenamefont {Xu}, \citenamefont
  {Duan},\ and\ \citenamefont {Xu}}]{gong23}%
  \BibitemOpen
  \bibfield  {author} {\bibinfo {author} {\bibfnamefont {X.}~\bibnamefont
  {Gong}}, \bibinfo {author} {\bibfnamefont {H.}~\bibnamefont {Li}}, \bibinfo
  {author} {\bibfnamefont {N.}~\bibnamefont {Zou}}, \bibinfo {author}
  {\bibfnamefont {R.}~\bibnamefont {Xu}}, \bibinfo {author} {\bibfnamefont
  {W.}~\bibnamefont {Duan}},\ and\ \bibinfo {author} {\bibfnamefont
  {Y.}~\bibnamefont {Xu}},\ }\bibfield  {title} {\bibinfo {title} {General
  framework for e(3)-equivariant neural network representation of density
  functional theory hamiltonian},\ }\href
  {https://doi.org/10.1038/s41467-023-38468-8} {\bibfield  {journal} {\bibinfo
  {journal} {Nature Communications}\ }\textbf {\bibinfo {volume} {14}},\
  \bibinfo {pages} {2848} (\bibinfo {year} {2023})}\BibitemShut {NoStop}%
\bibitem [{\citenamefont {Ma}\ \emph {et~al.}(2023)\citenamefont {Ma},
  \citenamefont {Chen}, \citenamefont {Deng}, \citenamefont {Tenenbaum},
  \citenamefont {Du}, \citenamefont {Gan},\ and\ \citenamefont
  {Matusik}}]{ma23}%
  \BibitemOpen
  \bibfield  {author} {\bibinfo {author} {\bibfnamefont {P.}~\bibnamefont
  {Ma}}, \bibinfo {author} {\bibfnamefont {P.~Y.}\ \bibnamefont {Chen}},
  \bibinfo {author} {\bibfnamefont {B.}~\bibnamefont {Deng}}, \bibinfo {author}
  {\bibfnamefont {J.~B.}\ \bibnamefont {Tenenbaum}}, \bibinfo {author}
  {\bibfnamefont {T.}~\bibnamefont {Du}}, \bibinfo {author} {\bibfnamefont
  {C.}~\bibnamefont {Gan}},\ and\ \bibinfo {author} {\bibfnamefont
  {W.}~\bibnamefont {Matusik}},\ }\bibfield  {title} {\bibinfo {title}
  {Learning neural constitutive laws from motion observations for generalizable
  pde dynamics},\ }in\ \href@noop {} {\emph {\bibinfo {booktitle} {Proceedings
  of the 40th International Conference on Machine Learning}}},\ \bibinfo
  {series and number} {ICML'23}\ (\bibinfo  {publisher} {JMLR.org},\ \bibinfo
  {year} {2023})\BibitemShut {NoStop}%
\bibitem [{\citenamefont {Marx}\ and\ \citenamefont {Hutter}(2009)}]{marx09}%
  \BibitemOpen
  \bibfield  {author} {\bibinfo {author} {\bibfnamefont {D.}~\bibnamefont
  {Marx}}\ and\ \bibinfo {author} {\bibfnamefont {J.}~\bibnamefont {Hutter}},\
  }\href@noop {} {\emph {\bibinfo {title} {Ab initio molecular dynamics: basic
  theory and advanced methods}}}\ (\bibinfo  {publisher} {Cambridge University
  Press},\ \bibinfo {year} {2009})\BibitemShut {NoStop}%
\bibitem [{\citenamefont {Cheng}\ \emph {et~al.}(2023)\citenamefont {Cheng},
  \citenamefont {Zhang}, \citenamefont {Nguyen}, \citenamefont {Azarfar},
  \citenamefont {Chern},\ and\ \citenamefont {Baek}}]{cheng23b}%
  \BibitemOpen
  \bibfield  {author} {\bibinfo {author} {\bibfnamefont {X.}~\bibnamefont
  {Cheng}}, \bibinfo {author} {\bibfnamefont {S.}~\bibnamefont {Zhang}},
  \bibinfo {author} {\bibfnamefont {P.~C.~H.}\ \bibnamefont {Nguyen}}, \bibinfo
  {author} {\bibfnamefont {S.}~\bibnamefont {Azarfar}}, \bibinfo {author}
  {\bibfnamefont {G.-W.}\ \bibnamefont {Chern}},\ and\ \bibinfo {author}
  {\bibfnamefont {S.~S.}\ \bibnamefont {Baek}},\ }\bibfield  {title} {\bibinfo
  {title} {Convolutional neural networks for large-scale dynamical modeling of
  itinerant magnets},\ }\href
  {https://doi.org/10.1103/PhysRevResearch.5.033188} {\bibfield  {journal}
  {\bibinfo  {journal} {Phys. Rev. Res.}\ }\textbf {\bibinfo {volume} {5}},\
  \bibinfo {pages} {033188} (\bibinfo {year} {2023})}\BibitemShut {NoStop}%
\bibitem [{\citenamefont {Karniadakis}\ \emph {et~al.}(2021)\citenamefont
  {Karniadakis}, \citenamefont {Kevrekidis}, \citenamefont {Lu}, \citenamefont
  {Perdikaris}, \citenamefont {Wang},\ and\ \citenamefont
  {Yang}}]{karniadakis21}%
  \BibitemOpen
  \bibfield  {author} {\bibinfo {author} {\bibfnamefont {G.~E.}\ \bibnamefont
  {Karniadakis}}, \bibinfo {author} {\bibfnamefont {I.~G.}\ \bibnamefont
  {Kevrekidis}}, \bibinfo {author} {\bibfnamefont {L.}~\bibnamefont {Lu}},
  \bibinfo {author} {\bibfnamefont {P.}~\bibnamefont {Perdikaris}}, \bibinfo
  {author} {\bibfnamefont {S.}~\bibnamefont {Wang}},\ and\ \bibinfo {author}
  {\bibfnamefont {L.}~\bibnamefont {Yang}},\ }\bibfield  {title} {\bibinfo
  {title} {Physics-informed machine learning},\ }\href
  {https://doi.org/10.1038/s42254-021-00314-5} {\bibfield  {journal} {\bibinfo
  {journal} {Nature Reviews Physics}\ }\textbf {\bibinfo {volume} {3}},\
  \bibinfo {pages} {422} (\bibinfo {year} {2021})}\BibitemShut {NoStop}%
\bibitem [{\citenamefont {Raissi}\ \emph {et~al.}(2019)\citenamefont {Raissi},
  \citenamefont {Perdikaris},\ and\ \citenamefont {Karniadakis}}]{raissi19}%
  \BibitemOpen
  \bibfield  {author} {\bibinfo {author} {\bibfnamefont {M.}~\bibnamefont
  {Raissi}}, \bibinfo {author} {\bibfnamefont {P.}~\bibnamefont {Perdikaris}},\
  and\ \bibinfo {author} {\bibfnamefont {G.}~\bibnamefont {Karniadakis}},\
  }\bibfield  {title} {\bibinfo {title} {Physics-informed neural networks: A
  deep learning framework for solving forward and inverse problems involving
  nonlinear partial differential equations},\ }\href
  {https://doi.org/https://doi.org/10.1016/j.jcp.2018.10.045} {\bibfield
  {journal} {\bibinfo  {journal} {Journal of Computational Physics}\ }\textbf
  {\bibinfo {volume} {378}},\ \bibinfo {pages} {686} (\bibinfo {year}
  {2019})}\BibitemShut {NoStop}%
\bibitem [{\citenamefont {Holstein}(1959)}]{holstein59}%
  \BibitemOpen
  \bibfield  {author} {\bibinfo {author} {\bibfnamefont {T.}~\bibnamefont
  {Holstein}},\ }\bibfield  {title} {\bibinfo {title} {Studies of polaron
  motion: Part i. the molecular-crystal model},\ }\href
  {https://doi.org/https://doi.org/10.1016/0003-4916(59)90002-8} {\bibfield
  {journal} {\bibinfo  {journal} {Annals of Physics}\ }\textbf {\bibinfo
  {volume} {8}},\ \bibinfo {pages} {325} (\bibinfo {year} {1959})}\BibitemShut
  {NoStop}%
\bibitem [{\citenamefont {Noack}\ \emph {et~al.}(1991)\citenamefont {Noack},
  \citenamefont {Scalapino},\ and\ \citenamefont {Scalettar}}]{Noack91}%
  \BibitemOpen
  \bibfield  {author} {\bibinfo {author} {\bibfnamefont {R.~M.}\ \bibnamefont
  {Noack}}, \bibinfo {author} {\bibfnamefont {D.~J.}\ \bibnamefont
  {Scalapino}},\ and\ \bibinfo {author} {\bibfnamefont {R.~T.}\ \bibnamefont
  {Scalettar}},\ }\bibfield  {title} {\bibinfo {title} {Charge-density-wave and
  pairing susceptibilities in a two-dimensional electron-phonon model},\ }\href
  {https://doi.org/10.1103/PhysRevLett.66.778} {\bibfield  {journal} {\bibinfo
  {journal} {Phys. Rev. Lett.}\ }\textbf {\bibinfo {volume} {66}},\ \bibinfo
  {pages} {778} (\bibinfo {year} {1991})}\BibitemShut {NoStop}%
\bibitem [{\citenamefont {Zhang}\ \emph {et~al.}(2019)\citenamefont {Zhang},
  \citenamefont {Chiu}, \citenamefont {Costa}, \citenamefont {Batrouni},\ and\
  \citenamefont {Scalettar}}]{Zhang19}%
  \BibitemOpen
  \bibfield  {author} {\bibinfo {author} {\bibfnamefont {Y.-X.}\ \bibnamefont
  {Zhang}}, \bibinfo {author} {\bibfnamefont {W.-T.}\ \bibnamefont {Chiu}},
  \bibinfo {author} {\bibfnamefont {N.~C.}\ \bibnamefont {Costa}}, \bibinfo
  {author} {\bibfnamefont {G.~G.}\ \bibnamefont {Batrouni}},\ and\ \bibinfo
  {author} {\bibfnamefont {R.~T.}\ \bibnamefont {Scalettar}},\ }\bibfield
  {title} {\bibinfo {title} {Charge order in the holstein model on a honeycomb
  lattice},\ }\href {https://doi.org/10.1103/PhysRevLett.122.077602} {\bibfield
   {journal} {\bibinfo  {journal} {Phys. Rev. Lett.}\ }\textbf {\bibinfo
  {volume} {122}},\ \bibinfo {pages} {077602} (\bibinfo {year}
  {2019})}\BibitemShut {NoStop}%
\bibitem [{\citenamefont {Chen}\ \emph {et~al.}(2019)\citenamefont {Chen},
  \citenamefont {Xu}, \citenamefont {Meng},\ and\ \citenamefont
  {Hohenadler}}]{Chen19}%
  \BibitemOpen
  \bibfield  {author} {\bibinfo {author} {\bibfnamefont {C.}~\bibnamefont
  {Chen}}, \bibinfo {author} {\bibfnamefont {X.~Y.}\ \bibnamefont {Xu}},
  \bibinfo {author} {\bibfnamefont {Z.~Y.}\ \bibnamefont {Meng}},\ and\
  \bibinfo {author} {\bibfnamefont {M.}~\bibnamefont {Hohenadler}},\ }\bibfield
   {title} {\bibinfo {title} {Charge-density-wave transitions of dirac fermions
  coupled to phonons},\ }\href {https://doi.org/10.1103/PhysRevLett.122.077601}
  {\bibfield  {journal} {\bibinfo  {journal} {Phys. Rev. Lett.}\ }\textbf
  {\bibinfo {volume} {122}},\ \bibinfo {pages} {077601} (\bibinfo {year}
  {2019})}\BibitemShut {NoStop}%
\bibitem [{\citenamefont {Esterlis}\ \emph {et~al.}(2019)\citenamefont
  {Esterlis}, \citenamefont {Kivelson},\ and\ \citenamefont
  {Scalapino}}]{Esterlis19}%
  \BibitemOpen
  \bibfield  {author} {\bibinfo {author} {\bibfnamefont {I.}~\bibnamefont
  {Esterlis}}, \bibinfo {author} {\bibfnamefont {S.~A.}\ \bibnamefont
  {Kivelson}},\ and\ \bibinfo {author} {\bibfnamefont {D.~J.}\ \bibnamefont
  {Scalapino}},\ }\bibfield  {title} {\bibinfo {title} {Pseudogap crossover in
  the electron-phonon system},\ }\href
  {https://doi.org/10.1103/PhysRevB.99.174516} {\bibfield  {journal} {\bibinfo
  {journal} {Phys. Rev. B}\ }\textbf {\bibinfo {volume} {99}},\ \bibinfo
  {pages} {174516} (\bibinfo {year} {2019})}\BibitemShut {NoStop}%
\bibitem [{\citenamefont {Li}\ \emph {et~al.}(2005)\citenamefont {Li},
  \citenamefont {Tully}, \citenamefont {Schlegel},\ and\ \citenamefont
  {Frisch}}]{li05x}%
  \BibitemOpen
  \bibfield  {author} {\bibinfo {author} {\bibfnamefont {X.}~\bibnamefont
  {Li}}, \bibinfo {author} {\bibfnamefont {J.~C.}\ \bibnamefont {Tully}},
  \bibinfo {author} {\bibfnamefont {H.~B.}\ \bibnamefont {Schlegel}},\ and\
  \bibinfo {author} {\bibfnamefont {M.~J.}\ \bibnamefont {Frisch}},\ }\bibfield
   {title} {\bibinfo {title} {Ab initio ehrenfest dynamics},\ }\href
  {https://doi.org/10.1063/1.2008258} {\bibfield  {journal} {\bibinfo
  {journal} {The Journal of Chemical Physics}\ }\textbf {\bibinfo {volume}
  {123}},\ \bibinfo {pages} {084106} (\bibinfo {year} {2005})}\BibitemShut
  {NoStop}%
\bibitem [{\citenamefont {Petrovic}\ \emph {et~al.}(2022)\citenamefont
  {Petrovic}, \citenamefont {Weber},\ and\ \citenamefont
  {Freericks}}]{Petrovic22}%
  \BibitemOpen
  \bibfield  {author} {\bibinfo {author} {\bibfnamefont {M.~D.}\ \bibnamefont
  {Petrovic}}, \bibinfo {author} {\bibfnamefont {M.}~\bibnamefont {Weber}},\
  and\ \bibinfo {author} {\bibfnamefont {J.~K.}\ \bibnamefont {Freericks}},\
  }\href@noop {} {\bibinfo {title} {Theoretical description of time-resolved
  photoemission in charge-density-wave materials out to long times}} (\bibinfo
  {year} {2022}),\ \Eprint {https://arxiv.org/abs/2203.11880} {arXiv:2203.11880
  [cond-mat.str-el]} \BibitemShut {NoStop}%
\bibitem [{\citenamefont {He}\ \emph {et~al.}(2016)\citenamefont {He},
  \citenamefont {Zhang}, \citenamefont {Ren},\ and\ \citenamefont
  {Sun}}]{he2016identitymappingsdeepresidual}%
  \BibitemOpen
  \bibfield  {author} {\bibinfo {author} {\bibfnamefont {K.}~\bibnamefont
  {He}}, \bibinfo {author} {\bibfnamefont {X.}~\bibnamefont {Zhang}}, \bibinfo
  {author} {\bibfnamefont {S.}~\bibnamefont {Ren}},\ and\ \bibinfo {author}
  {\bibfnamefont {J.}~\bibnamefont {Sun}},\ }\href
  {https://arxiv.org/abs/1603.05027} {\bibinfo {title} {Identity mappings in
  deep residual networks}} (\bibinfo {year} {2016}),\ \Eprint
  {https://arxiv.org/abs/1603.05027} {arXiv:1603.05027 [cs.CV]} \BibitemShut
  {NoStop}%
\bibitem [{\citenamefont {Ioffe}\ and\ \citenamefont
  {Szegedy}(2015)}]{ioffe2015batchnormalizationacceleratingdeep}%
  \BibitemOpen
  \bibfield  {author} {\bibinfo {author} {\bibfnamefont {S.}~\bibnamefont
  {Ioffe}}\ and\ \bibinfo {author} {\bibfnamefont {C.}~\bibnamefont
  {Szegedy}},\ }\href {https://arxiv.org/abs/1502.03167} {\bibinfo {title}
  {Batch normalization: Accelerating deep network training by reducing internal
  covariate shift}} (\bibinfo {year} {2015}),\ \Eprint
  {https://arxiv.org/abs/1502.03167} {arXiv:1502.03167 [cs.LG]} \BibitemShut
  {NoStop}%
\bibitem [{\citenamefont {Ba}\ \emph {et~al.}(2016)\citenamefont {Ba},
  \citenamefont {Kiros},\ and\ \citenamefont
  {Hinton}}]{ba2016layernormalization}%
  \BibitemOpen
  \bibfield  {author} {\bibinfo {author} {\bibfnamefont {J.~L.}\ \bibnamefont
  {Ba}}, \bibinfo {author} {\bibfnamefont {J.~R.}\ \bibnamefont {Kiros}},\ and\
  \bibinfo {author} {\bibfnamefont {G.~E.}\ \bibnamefont {Hinton}},\ }\href
  {https://arxiv.org/abs/1607.06450} {\bibinfo {title} {Layer normalization}}
  (\bibinfo {year} {2016}),\ \Eprint {https://arxiv.org/abs/1607.06450}
  {arXiv:1607.06450 [stat.ML]} \BibitemShut {NoStop}%
\bibitem [{\citenamefont {Tompson}\ \emph {et~al.}(2014)\citenamefont
  {Tompson}, \citenamefont {Goroshin}, \citenamefont {Jain}, \citenamefont
  {LeCun},\ and\ \citenamefont {Bregler}}]{tompson14}%
  \BibitemOpen
  \bibfield  {author} {\bibinfo {author} {\bibfnamefont {J.}~\bibnamefont
  {Tompson}}, \bibinfo {author} {\bibfnamefont {R.}~\bibnamefont {Goroshin}},
  \bibinfo {author} {\bibfnamefont {A.}~\bibnamefont {Jain}}, \bibinfo {author}
  {\bibfnamefont {Y.}~\bibnamefont {LeCun}},\ and\ \bibinfo {author}
  {\bibfnamefont {C.}~\bibnamefont {Bregler}},\ }\bibfield  {title} {\bibinfo
  {title} {Efficient object localization using convolutional networks},\ }\href
  {https://api.semanticscholar.org/CorpusID:206592615} {\bibfield  {journal}
  {\bibinfo  {journal} {2015 IEEE Conference on Computer Vision and Pattern
  Recognition (CVPR)}\ ,\ \bibinfo {pages} {648}} (\bibinfo {year}
  {2014})}\BibitemShut {NoStop}%
\bibitem [{\citenamefont {Goodfellow}\ \emph
  {et~al.}(2016{\natexlab{b}})\citenamefont {Goodfellow}, \citenamefont
  {Bengio},\ and\ \citenamefont {Courville}}]{Goodfellow-et-al-2016}%
  \BibitemOpen
  \bibfield  {author} {\bibinfo {author} {\bibfnamefont {I.}~\bibnamefont
  {Goodfellow}}, \bibinfo {author} {\bibfnamefont {Y.}~\bibnamefont {Bengio}},\
  and\ \bibinfo {author} {\bibfnamefont {A.}~\bibnamefont {Courville}},\
  }\href@noop {} {\emph {\bibinfo {title} {Deep Learning}}}\ (\bibinfo
  {publisher} {MIT Press},\ \bibinfo {year} {2016})\ \bibinfo {note}
  {\url{http://www.deeplearningbook.org}}\BibitemShut {NoStop}%
\bibitem [{\citenamefont {Loshchilov}\ and\ \citenamefont
  {Hutter}(2017{\natexlab{a}})}]{loshchilov17}%
  \BibitemOpen
  \bibfield  {author} {\bibinfo {author} {\bibfnamefont {I.}~\bibnamefont
  {Loshchilov}}\ and\ \bibinfo {author} {\bibfnamefont {F.}~\bibnamefont
  {Hutter}},\ }\bibfield  {title} {\bibinfo {title} {Decoupled weight decay
  regularization},\ }in\ \href
  {https://api.semanticscholar.org/CorpusID:53592270} {\emph {\bibinfo
  {booktitle} {International Conference on Learning Representations}}}\
  (\bibinfo {year} {2017})\BibitemShut {NoStop}%
\bibitem [{\citenamefont {Kingma}\ and\ \citenamefont {Ba}(2017)}]{kingma17}%
  \BibitemOpen
  \bibfield  {author} {\bibinfo {author} {\bibfnamefont {D.~P.}\ \bibnamefont
  {Kingma}}\ and\ \bibinfo {author} {\bibfnamefont {J.}~\bibnamefont {Ba}},\
  }\href {https://arxiv.org/abs/1412.6980} {\bibinfo {title} {Adam: A method
  for stochastic optimization}} (\bibinfo {year} {2017}),\ \Eprint
  {https://arxiv.org/abs/1412.6980} {arXiv:1412.6980 [cs.LG]} \BibitemShut
  {NoStop}%
\bibitem [{\citenamefont {Pascanu}\ \emph {et~al.}(2013)\citenamefont
  {Pascanu}, \citenamefont {Mikolov},\ and\ \citenamefont
  {Bengio}}]{pascanu2013difficultytrainingrecurrentneural}%
  \BibitemOpen
  \bibfield  {author} {\bibinfo {author} {\bibfnamefont {R.}~\bibnamefont
  {Pascanu}}, \bibinfo {author} {\bibfnamefont {T.}~\bibnamefont {Mikolov}},\
  and\ \bibinfo {author} {\bibfnamefont {Y.}~\bibnamefont {Bengio}},\ }\href
  {https://arxiv.org/abs/1211.5063} {\bibinfo {title} {On the difficulty of
  training recurrent neural networks}} (\bibinfo {year} {2013}),\ \Eprint
  {https://arxiv.org/abs/1211.5063} {arXiv:1211.5063 [cs.LG]} \BibitemShut
  {NoStop}%
\bibitem [{\citenamefont {Bengio}\ \emph {et~al.}(2009)\citenamefont {Bengio},
  \citenamefont {Louradour}, \citenamefont {Collobert},\ and\ \citenamefont
  {Weston}}]{bengio2009curriculum}%
  \BibitemOpen
  \bibfield  {author} {\bibinfo {author} {\bibfnamefont {Y.}~\bibnamefont
  {Bengio}}, \bibinfo {author} {\bibfnamefont {J.}~\bibnamefont {Louradour}},
  \bibinfo {author} {\bibfnamefont {R.}~\bibnamefont {Collobert}},\ and\
  \bibinfo {author} {\bibfnamefont {J.}~\bibnamefont {Weston}},\ }\bibfield
  {title} {\bibinfo {title} {Curriculum learning}\ }(\bibinfo  {publisher}
  {Association for Computing Machinery},\ \bibinfo {address} {New York, NY,
  USA},\ \bibinfo {year} {2009})\ p.\ \bibinfo {pages} {41–48}\BibitemShut
  {NoStop}%
\bibitem [{\citenamefont {Loshchilov}\ and\ \citenamefont
  {Hutter}(2017{\natexlab{b}})}]{loshchilov2017sgdrstochasticgradientdescent}%
  \BibitemOpen
  \bibfield  {author} {\bibinfo {author} {\bibfnamefont {I.}~\bibnamefont
  {Loshchilov}}\ and\ \bibinfo {author} {\bibfnamefont {F.}~\bibnamefont
  {Hutter}},\ }\href {https://arxiv.org/abs/1608.03983} {\bibinfo {title}
  {Sgdr: Stochastic gradient descent with warm restarts}} (\bibinfo {year}
  {2017}{\natexlab{b}}),\ \Eprint {https://arxiv.org/abs/1608.03983}
  {arXiv:1608.03983 [cs.LG]} \BibitemShut {NoStop}%
\bibitem [{\citenamefont {Horsfield}\ \emph {et~al.}(2004)\citenamefont
  {Horsfield}, \citenamefont {Bowler}, \citenamefont {Fisher}, \citenamefont
  {Todorov},\ and\ \citenamefont {Sánchez}}]{horsfield04}%
  \BibitemOpen
  \bibfield  {author} {\bibinfo {author} {\bibfnamefont {A.~P.}\ \bibnamefont
  {Horsfield}}, \bibinfo {author} {\bibfnamefont {D.~R.}\ \bibnamefont
  {Bowler}}, \bibinfo {author} {\bibfnamefont {A.~J.}\ \bibnamefont {Fisher}},
  \bibinfo {author} {\bibfnamefont {T.~N.}\ \bibnamefont {Todorov}},\ and\
  \bibinfo {author} {\bibfnamefont {C.~G.}\ \bibnamefont {Sánchez}},\
  }\bibfield  {title} {\bibinfo {title} {Beyond ehrenfest: correlated
  non-adiabatic molecular dynamics},\ }\href
  {https://doi.org/10.1088/0953-8984/16/46/012} {\bibfield  {journal} {\bibinfo
   {journal} {Journal of Physics: Condensed Matter}\ }\textbf {\bibinfo
  {volume} {16}},\ \bibinfo {pages} {8251} (\bibinfo {year}
  {2004})}\BibitemShut {NoStop}%
\bibitem [{\citenamefont {Tapavicza}\ \emph {et~al.}(2013)\citenamefont
  {Tapavicza}, \citenamefont {Bellchambers}, \citenamefont {Vincent},\ and\
  \citenamefont {Furche}}]{tapavicza13}%
  \BibitemOpen
  \bibfield  {author} {\bibinfo {author} {\bibfnamefont {E.}~\bibnamefont
  {Tapavicza}}, \bibinfo {author} {\bibfnamefont {G.~D.}\ \bibnamefont
  {Bellchambers}}, \bibinfo {author} {\bibfnamefont {J.~C.}\ \bibnamefont
  {Vincent}},\ and\ \bibinfo {author} {\bibfnamefont {F.}~\bibnamefont
  {Furche}},\ }\bibfield  {title} {\bibinfo {title} {Ab initio non-adiabatic
  molecular dynamics},\ }\href {https://doi.org/10.1039/C3CP51514A} {\bibfield
  {journal} {\bibinfo  {journal} {Phys. Chem. Chem. Phys.}\ }\textbf {\bibinfo
  {volume} {15}},\ \bibinfo {pages} {18336} (\bibinfo {year}
  {2013})}\BibitemShut {NoStop}%
\bibitem [{\citenamefont {Jain}\ and\ \citenamefont {Sindhu}(2022)}]{jain22}%
  \BibitemOpen
  \bibfield  {author} {\bibinfo {author} {\bibfnamefont {A.}~\bibnamefont
  {Jain}}\ and\ \bibinfo {author} {\bibfnamefont {A.}~\bibnamefont {Sindhu}},\
  }\bibfield  {title} {\bibinfo {title} {Pedagogical overview of the fewest
  switches surface hopping method},\ }\href
  {https://doi.org/10.1021/acsomega.2c04843} {\bibfield  {journal} {\bibinfo
  {journal} {ACS Omega}\ }\textbf {\bibinfo {volume} {7}},\ \bibinfo {pages}
  {45810} (\bibinfo {year} {2022})}\BibitemShut {NoStop}%
\end{thebibliography}%

\end{document}